\documentclass[]{aa}  

\usepackage{graphicx}
\usepackage{txfonts}
\usepackage{amsmath}	% Advanced maths commands
\usepackage{booktabs}
\usepackage{hyperref}
\usepackage{color}

\usepackage{lipsum}
\usepackage{subcaption}         % necessary for continued figures, example in section 3 and appendix
\usepackage{lscape}             % to rotate a single page table, example in appendix. For landscape tables, see the longtable examples.
\usepackage{placeins}           % useful with \FloatBarrier, to keep onecolumn floats from drifting to the next section

\defcitealias{jadhav_2021}{JS21}
\defcitealias{rain_2021}{R21}
\defcitealias{cordoni_2018}{C18}
\defcitealias{dattatrey_2023}{D23}

\begin{document}

   \title{Beyond the main sequence: Binary evolution pathways to blue stragglers in the Gaia era}

   \subtitle{I. Galactic open and globular clusters}

   \author{Francisco F. Carrasco-Varela\inst{1}
          \and
          Prasanta K. Nayak\inst{1}\fnmsep\thanks{ Corresponding author:  ffcarrasco@uc.cl; pnayak@astro.puc.cl; nayakphy@gmail.com}\fnmsep\thanks{CATA Research Fellow}
          \and
          Thomas H. Puzia\inst{1}
          }

   \institute{Institute of Astrophysics, Pontificia Universidad Cat{\'o}lica de Chile, Avenida Vicu{\'n}a Mackenna 4860, 7820436, Macul, Santiago, Chile\\
              % \email{nayakphy@gmail.com}
             }

  \date{Received 10 September 2024; accepted Accepted 26 March 2025}

% \abstract{}{}{}{}{}
% 5 {} token are mandatory
 
  \abstract
   {The study of blue straggler stars (BSSs) provides insight into the mechanisms of stellar mass exchange during binary stellar evolution and the complex gravitational interactions within dense stellar systems. In combination, they enhance our understanding of the possible life cycles of stars and the evolutionary pathways of star clusters.} 
   {We study the populations of BSSs in 41 globular clusters (GCs) and 42 open clusters (OCs) based on photometry, proper motion, and parallax from the Gaia Data Release 3 (DR3). We confirm their cluster membership.~We find a total of 4399 BSSs: All GCs show BSSs (3965 or $\sim\!90\%$ of the sample), whereas only 42 out of 129 studied OCs show BSSs (434 or $\sim\!10\%$ of the sample). Clusters younger than $\sim\!500$\,Myr do not host any BSSs.}
   {We derived their astrophysical parameters such as effective temperature, surface gravity, and mass based on color-temperature relations, isochrone models, and Gaia DR3 spectroscopy (if available). We found values for $T_{\rm eff}=(6800 \pm 585) \ {\rm K}$ and $(7570 \pm 1400) \ {\rm K}$ and an average mass of $\langle M_{\rm BSS} \rangle = (1.02 \pm 0.1) \ {M_\odot}$ and $\langle M_{\rm BSS} \rangle = (1.75 \pm 0.45) \ {M_\odot}$ for GCs and OCs, respectively. We finally computed the difference of the BSS mass and the main-sequence turn-off (MSTO) mass of its respective cluster, normalized by the MSTO mass, for every identified BSS.}
   {Based on this parameter and on the BSSs ages derived from isochrone models, we find that i) GC BSSs that are most likely to be formed through collisions show a boost for ages $\sim\!1\!-\!2$\,Gyr. This agrees with the ages for core-collapse events in GCs that were reported in previous studies. We also find ii) a double sequence for GC BSSs that might indicate a pre- or post-merger or close-binary scenario.}
   {}


   % blue stragglers – stars: fundamental parameters – globular clusters: general – open clusters and associations: general

   \maketitle

%________________________________________________________________

\section{Introduction}\label{intro}
Blue straggler stars (BSSs) are stars that appear to be younger and more massive than the bulk of other stars in their host stellar population. They are characterized by their location in the color-magnitude diagram (CMD), where they appear hotter and bluer than their main-sequence (MS) neighbors.

Blue straggler stars were first identified by \cite{sandage_1953} through his observations of the globular cluster (GC) Messier\,3 (M3), and some years later, they were reported in the open cluster (OC) NGC\,7786 by \citet{burbidge_1958}. Ever since, their presence has been noted in distinct structures: as field stars \citep{preston_2000, jofre_2016}, in OB associations \citep{mathys_1987}, in the Galactic bulge \citep{clarkson_2011}, and in dwarf galaxies \citep{da_costa_1984, mapelli_2009}.

Blue straggler stars challenge the standard theory of stellar evolution, which states that the more massive a star, the faster it evolves. For this reason, these stars should have evolved off the MS branch a long time ago, but they still populate this branch as an apparent extension of the MS in a CMD. Three scenarios have been proposed to explain their formation: i) Mass transfer (MT) in binary systems \citep{mccrea_1964}, where a star receives material from its companion. This extends its life. ii) Stars formed through collisions or mergers \citep{hills_1976, lombardi_2002}. iii) In hierarchical triple systems, magnetized winds or stellar evolution cause the loss of angular momentum in a multiple system, which leads to collisions (or MT) between the stars that conform the system \citep{perets_2009, naoz_2014}. It has been shown that the BSS properties scale with different astrophysical parameters of their parent cluster \citep{knigge_2009, leigh_sills_2011, jadhav_2021}, which might help us in principle to constrain which of the formation pathways dominates. It is important to mention, however, that these formation mechanisms most likely coexist, that is, they operate simultaneously \citep{mapelli_2006, leigh_sills_2011}.

\citet{ferraro_2009} found a double sequence of BSSs in the GC M30: a blue and a red BSS sequence, which were both roughly aligned with the zero-age MS (ZAMS). The authors argued that the distinction between these two BSS sequences may be due to their origin: collisional BSSs (Col-BSSs), which are deemed more massive and hotter and therefore appear bluer in the CMD, whereas MT-BSSs would be less massive, mildly cooler than Col-BSSs, and would therefore appear as a redder BSS sequence.~This is not an isolated case. Double/multiple sequences of BSSs have been reported for more GCs, such as NGC\,362 \citep{dalessandro_2013}, NGC\,1261 \citep{simunovic_2014}, M\,15 \citep{beccari_2019}, and NGC\,6256 \citep{double_BSS_NGC6256}. The explanation given for all these cases leads to a core-collapse scenario of the host GC in which the central interactions between stars increase, resulting in the formation of BSSs. Therefore, a double/multiple BSS sequence could be a vestige of this event. It also has been found that some BSSs show a hot companion \citep{gosnell_2015, sindhu_2019}, which provides supporting evidence for the MT formation scenario in action. Linking these concepts, \citet[][hereafter \citetalias{dattatrey_2023}]{dattatrey_2023} have analyzed the spectral energy distribution (SED) for multiple BSSs in the GC NGC\,362.~They demonstrated that BSSs with and without a hot companion lie on the red and blue BSS sequence, respectively, which agrees with the explanations provided by \citet{dalessandro_2013} for the same GC.
\noindent
More recently, a double BSS sequence has also been detected in the OC Berkeley\,17 by \citet{rao_2023b}. Given the lack of core collapse in OCs and the accordingly considerable attenuation of the Col-BSS formation channel, the authors argued that this double BSS sequence might be due to a significant difference in BSS rotational velocities, the presence of multiple stellar populations (MSP), or to extinction effects \citep[for more details see][and references therein]{rao_2023b}.

The arrival of the Gaia mission \citep{gaia_2016} with data for almost 1.7 billion objects has opened new opportunities to explore the kinematics of our  Galaxy and has helped us to identify and confirm the cluster membership of stars in GCs \citep{vasiliev_2021} and OCs \citep{cantat_gaudin_2018, dias_2021}. Gaia faces limitations in overcrowded regions (large errors for its astrometric parameters because it cannot reliably resolve sources), however, such as GC centers. Nevertheless, BSS studies in crowded regions are covered by HST\footnote{Hubble Space Telescope} data, which can lead to high-accuracy measurements in the densest regions \citep[e.g.][]{baldwin_2016}, and because HST features a broader SED coverage than Gaia, BSSs can also be easily identified through the so-called UV-route \citep[and references therein]{ferraro_1997, raso_2017, ferraro_2018}. 

It is particularly interesting to analyze the radial distribution of BSSs, especially in the inner regions of clusters.~Since BSSs are more massive compared with other stars in the same cluster, they tend to undergo efficient mass segregation, that is, they sink to the inner parts of the cluster \citep{contreras_ramos_2012}. It is therefore expected that these peculiar stars can be used to measure the dynamical state of a cluster, that is, they can be used as dynamical clocks in GCs \citep{ferraro_2012, ferraro_2023} and OCs \citep{rao_2021, rao_2023a}.  

Gaia Data Release 3 \citep[hereafter DR3]{gaia_2023} for OCs does not suffer from overcrowding, even in the core of OCs.~For example, using Gaia Data Release 2 \citep[hereafter DR2]{gaia_2018}, \citet[hereafter \citetalias{rain_2021}]{rain_2021} created a catalog for 408 BSSs in different OCs, while \citet[][hereafter \citetalias{jadhav_2021}]{jadhav_2021} have studied the scaling relations of BSSs and their parent OCs. Gaia DR2 has also opened new opportunities for studying GCs and their kinematics \citep[e.g.,][]{vasiliev_2021}. Several HST-based studies have previously characterized BSSs in GCs \citep[][]{simunovic_2016}. 
The limitation of HST-based studies was, however, that they can only cover the inner regions of clusters, that is, from the center up to about the half-light radius, $r_{\rm hl}$. For example, data from the HST UV Globular Cluster Survey \citep[also known as ``HUGS'']{piotto_2015, nardiello_2018} cover this region and do not include the outskirts of the target GCs. Therefore, combining Gaia DR3 with HST will help us to analyze the radial distribution of BSSs in GCs from the core to the outer radii and to determine whether the GCs are dynamically settled. In the case of OCs, Gaia DR3 alone can be used to study their dynamical evolution. Hence, a homogeneous catalog of BSSs in GCs and OCs is necessary to understand their formation channels and to examine the dynamical evolution of their host clusters.

We use Gaia DR3 to create a new homogeneous BSSs catalog of GCs and OCs that differs from previous catalogs \citep{rain_2021, simunovic_2016}. This work includes BSSs that were previously not covered due to the radial coverage limitations of prior work. This study also provides various newly derived astrophysical parameters of BSSs from the color--temperature relation, isochrone models, and DR3 spectra. 

This paper is organized as follows. In Section~\ref{sec:data_selection} we describe the selection of cluster members and the selection of BSSs. We introduce the different methods we applied to obtain different astrophysical parameters for the BSS cluster members we identified. In Section~\ref{sec:results} we present our results for the total number of identified BSSs and their fraction with respect to their parent cluster MS stars, as well as the distributions for the derived astrophysical BSS parameters. We analyze the results and compare them with previous results in the literature, and we discuss the implications. We finally discuss possible uses and extensions for the identified BSS sample.~In Section~\ref{sec:conclusion} we summarize the main results of this work.
%__________________________________________________________________

\section{Data selection}\label{sec:data_selection}

To select the clusters studied in this work, we first select GCs and OCs that have a significant number of members based on previous studies.  
We select all GCs from the catalog by \citet{vasiliev_2021} with at least 1000 identified member stars and all OCs from the catalog by \citet{dias_2021} with at least 350 identified member stars. We find 41 GCs and 129 OCs following the above criteria.

\subsection{Star cluster membership} \label{subsect:data_sel}

To identify cluster members, we apply a technique developed and used by \citet[][hereafter \citetalias{cordoni_2018}]{cordoni_2018}, with a minor change for OCs (see step 2.), which has also been used by \citet{cordoni_2020} for GCs. Our procedure is summarized as follows:

\begin{enumerate}
    \item First, we select data from Gaia DR3 based on coordinates provided by \citet{vasiliev_2021} for GCs and \citet{dias_2021} for OCs. As the data query radius, we use the \texttt{Rscale} value --the scale radius of Gaia-detected cluster members--, plus  $\sim 2~{\rm arcmin}$, from \citet{vasiliev_2021} for GCs. For OCs we select stars inside $\sim 5 \times r_{50}$ --the radius containing half of the identified members from \citet{dias_2021}--, with a maximum apparent radius of $80~{\rm arcmin}$.
    \item We only keep data with G magnitudes in the range between $\sim\!10$ and $19.5$, with an error in proper motion (in both components along RA and Dec) lower than $0.35 \ {\rm mas} \cdot {\rm yr}^{-1}$ and the renormalized unit weight error (\texttt{ruwe}) lower than 1.4.
    \item We select stars inside an ellipse (note: \citetalias{cordoni_2018} uses a circle) in a vector point diagram (VPD), whose center is based on proper motion coordinates provided by \citet{vasiliev_2021} for GCs and \citet{dias_2021} for OCs. We slightly vary the minor and major axis of this ellipse, as well as its inclination angle with respect to the y-axis in the VPD, and we keep those parameters that maximize the number of stars within the ellipse.
    \item We compute the parameter $\mu_{\rm R}$, defined by \citetalias{cordoni_2018}, as:
    \begin{equation}\label{eq:mu_R}
        \mu_{\rm R} = \sqrt{ \big( \mu_{\alpha}\cos \delta - \langle \mu_{\alpha} \cos \delta \rangle \big)^2 + \big( \mu_\delta - \langle \mu_\delta \rangle \big)^2}
    \end{equation}
    where $\langle \mu_{\alpha} \cos \delta \rangle$ and $\langle \mu_\delta \rangle$ are the median of the proper motion along RA and DEC components, respectively, for data within the chosen ellipse. In short, $\mu_{\rm R}$ is the distance of every star in the VPD-space to the center of the ellipse.
    \item We plot the G magnitude versus $\mu_{\rm R}$ and split the magnitude component into $20$ bins with bin size of $0.5$ mag. In every bin, we compute the median values for ${\rm G}$ mag and the $\mu_{\rm R}$ parameter, and the rms ($\sigma$) for $\mu_{\rm R}$. We then create points with coordinates G median (vertical axis) and $\mu_{\rm R} \pm 3 \sigma_{\mu_{\rm R}}$ (horizontal axis). We interpolate all the points created in this way and we keep all the values within $\pm 3 \sigma_{\mu_{\rm R}}$ under this interpolation.
    \item We repeat the same procedure as the previous step, but with the parallax ($\varpi$) instead of $\mu_{\rm R}$. For every bin, we create two boundaries with coordinate points-based median ${\rm G}$ mag (vertical axis) and median $\varpi \pm 3 \sigma_{\varpi}$ (horizontal axis), interpolate them, and keep the data between these $\pm 3 \sigma_{\varpi}$ boundaries.
\end{enumerate}

\begin{figure}
 	\includegraphics[width=\columnwidth]{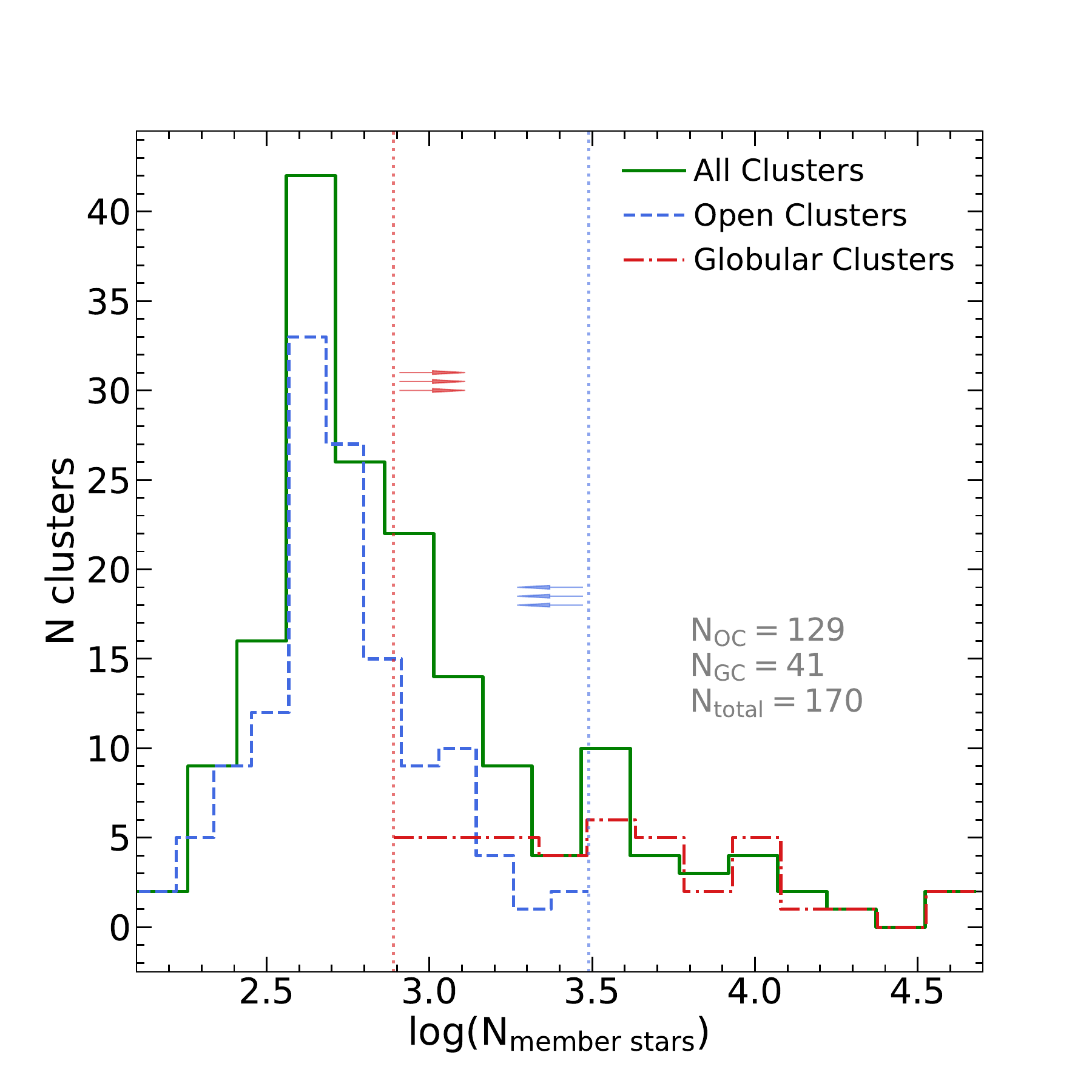}
    \caption{Histogram for the identified member stars in our sample GCs and OCs (see Sect.~\ref{subsect:data_sel} for details). The distribution for GCs is shown as a dash--dotted red line, for OCs as a dashed blue line, and for the composite sample (GCs+OCs) as a solid green line.~The dotted vertical red line with arrows pointing to the right denotes the minimum number of member stars for GCs, and the dotted vertical blue line with arrows pointing to the left indicates the maximum number of identified member stars in OCs.}
    \label{fig:cluster_all_members}
\end{figure}

We iterate this procedure a total of 3 times starting from step 3. For a more detailed description of this procedure, we encourage the reader to view section~2 and figure~1 from \citetalias{cordoni_2018} and \citet{cordoni_2020}. Figure~\ref{fig:cluster_all_members} shows the distribution of the member stars in our GC and OC sample. We also compared our cluster members with the literature \citep{dias_2021, vasiliev_2021, Hunt_2023_OCs, Cantat-Gaudin_2020_OCs} and found a very good match with them. The small discrepancy in the crossmatch is found mostly towards the fainter magnitudes (G$>$18). The details of this crossmatch are described in \autoref{membership_validation}.

\subsection{Selection of BSSs }\label{sec:bss_selection_criteria}

We created a CMD ridge-line for each cluster using the median color and magnitude in $\sim\!40$ bins in the ${\rm G}_{\rm BP}$ versus ${\rm G}_{\rm BP} - {\rm G}_{\rm RP}$ CMD and select MS, subgiant branch (SGB) and red giant branch (RGB) stars.~We prefer ${\rm G}_{\rm BP}$ over the ${\rm G}$ magnitude in the CMD since it is expected that BSSs will appear brighter at bluer wavelengths across the Gaia SED coverage \citep[see, e.g.,][]{raso_2017}.~The difference of using ${\rm G}$ or ${\rm G}_{\rm BP}$ does not change our final sample size, however, because ${\rm G}$ covers approximately $330$--$1040 \ {\rm nm}$, while ${\rm G}_{\rm BP}$ covers $330$--$640 \ {\rm nm}$ \citep{solano_2012, solano_2020}, and given the lack of significant near-IR emission of BSS star the wider coverage of the ${\rm G}$ filter mainly adds noise and not useful flux information for BSSs \citepalias[see e.g.][]{dattatrey_2023}. 

Next, we use PARSEC isochrones \citep{bressan_2012} that best fit these ridge lines, adopting the stellar population parameters from the literature: For GCs, distances were taken from \citet{baumgardt_2021}, ages from \citet{baumgardt_2023} and extinctions from \citet{harris_2010}. All parameters for OCs were taken from \citet{dias_2021}, except for two OCs: for Pismin 3 we use parameters from \citet{bisht_2022} and for Trumpler 23 we use parameters from \citet{cantat_gaudin_2020}. Starting from these literature values, we also generate isochrones with slight input parameter variations to explore the best-fit isochrone morphologies: $\pm 0.5~{\rm dex}$ in both $\log({\rm age})$ and $[{\rm M}/{\rm H}]$ in steps of $0.05~{\rm dex}$. 
~With the color excess--extinction relation ${\rm A}_{\rm V} = 3.1 \times {\rm E}({\rm B}-{\rm V})$ \citep[see][]{schultz_1975, cardelli_1989} converted to Gaia extinction coefficients using the relations ${\rm A}_{\rm V} = 1.06794 \times {\rm A}_{\rm BP}$ and ${\rm A}_{\rm V} = 0.65199 \times {\rm A}_{\rm RP}$ \citep{marigo_2008, evans_2018}.~We consider the PARSEC isochrone that best fits our ridge line for each cluster using a least-square fit.

\begin{figure}
 	\includegraphics[width=\columnwidth]{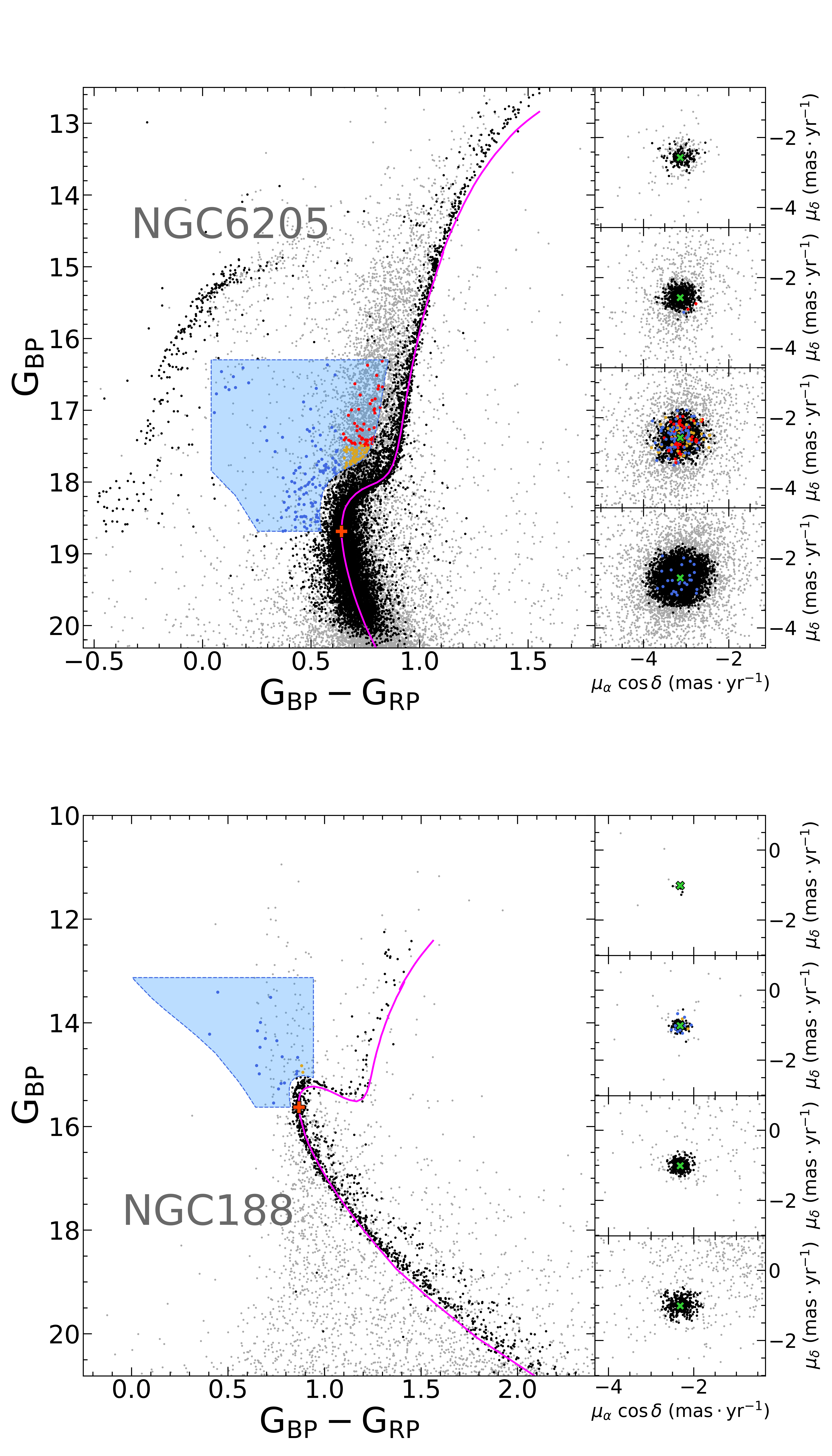}
    \caption{\textit{Top:} CMD of the GC NGC\,6205 using Gaia DR3. Stars that were discarded as cluster members using our selection procedure are shown as gray dots, while cluster members are plotted as black dots. The blue shaded area represents our selection area for BSSs, YSSs and RSSs (see Sect.~\ref{sec:bss_selection_criteria}) which are shown as blue, yellow and red dots, respectively. The magenta solid curve represents the best-fitting PARSEC isochrone, with the \texttt{+} symbol indicating the MSTO. The right panels show the VPDs for the stars in the respective magnitude range from the left panel. Green crosses represent the median of confirmed members for both components in the VPD. \textit{Bottom:} Same as the top figure, but for the OC NGC\,188. We added a red border in the selection area for OCs to avoid AGB/RGB stars (see text for details).}
    \label{fig:bss_selection_example}
\end{figure}

With the best-fitting isochrones, we divide the data below the MS turn-off (MSTO) point --determined by the infinite slope of the best-fitting isochrone-- into 10 bins and record the bin with the largest number of stars. For this bin, we compute its root mean square (rms) along the color axis and call it $\sigma_{\rm color}$. We repeat the procedure, but now for SGB stars, and measure the rms along the magnitude axis to obtain $\sigma_{G_{\rm BP}}$. To properly select BSSs, we create a region where we shift the best-fitting isochrone $3\sigma_{\rm color}$ to bluer colors and $4\sigma_{\rm BP}$ to brighter magnitudes, and adopt an upper cut in magnitude $2$ mag above the MSTO for GCs, and 2.5 mags for OCs \citep{chen_2009, leigh_sills_2011}. 

Finally, we recreate a zero-age main sequence (ZAMS) keeping the stellar population parameters for the best-fitting isochrone and shifting it $6\sigma_{\rm color}$ to bluer colors, i.e. twice the quantity we shifted the best-fitting isochrone before.~The only 2 differences in the BSS selection between GCs and OCs are in the color selection: i) for GCs we set a cut in the color up to 0.5 mag to colors bluer than the MSTO to avoid selecting evolved horizontal branch (HB) stars \citep[see, e.g.,][]{culpan_2021}. We notice that with this criteria, the selected BSS candidates are found to be located at least five sigma away from the BHB ridge line for all GCs. We did not apply this criterion to OCs, however, because they have no BHB populations.   
ii) For OCs, we applied a cut in color to ${\rm MSTO}_{\rm color} - 3\sigma_{\rm color}$ mag to avoid selecting RGB stars for younger clusters. 

Finally, we classify stars within this selection region into 3 types: i) if they are bluer than the MSTO they are classified as BSSs; ii) if they are redder than the MSTO and are in the fainter half of the selection region they are classified as yellow straggler stars (YSSs), i.e., evolved BSSs; and iii) if they are redder than the MSTO and lie in the brighter-half of the selection region they are classified as red straggler stars (RSSs), or evolved YSSs. In this work, we will mainly focus on stars classified as BSSs.

A visual representation of our selection criteria can be appreciated in Figure~\ref{fig:bss_selection_example} for two representative clusters, GC NGC\,6205 and OC NGC\,188, where we display the sample selection results along with their respective VPDs.

We detect 4399 BSSs in our cluster sample. All GCs show BSSs, with a total of 3965 detected BSSs ($\sim\!90\%$ of the sample). On the other hand, only 42 out of 129 selected OCs in our sample show BSSs, with a total of 434 BSSs ($\sim\!10\%$ of the sample). Comparing our BSS members with the literature, we find that more than 90\% of all BSSs are also classified as members in previous studies.

%________________________________________________________________

\begin{figure*}[ht!]
	\includegraphics[width=\textwidth]{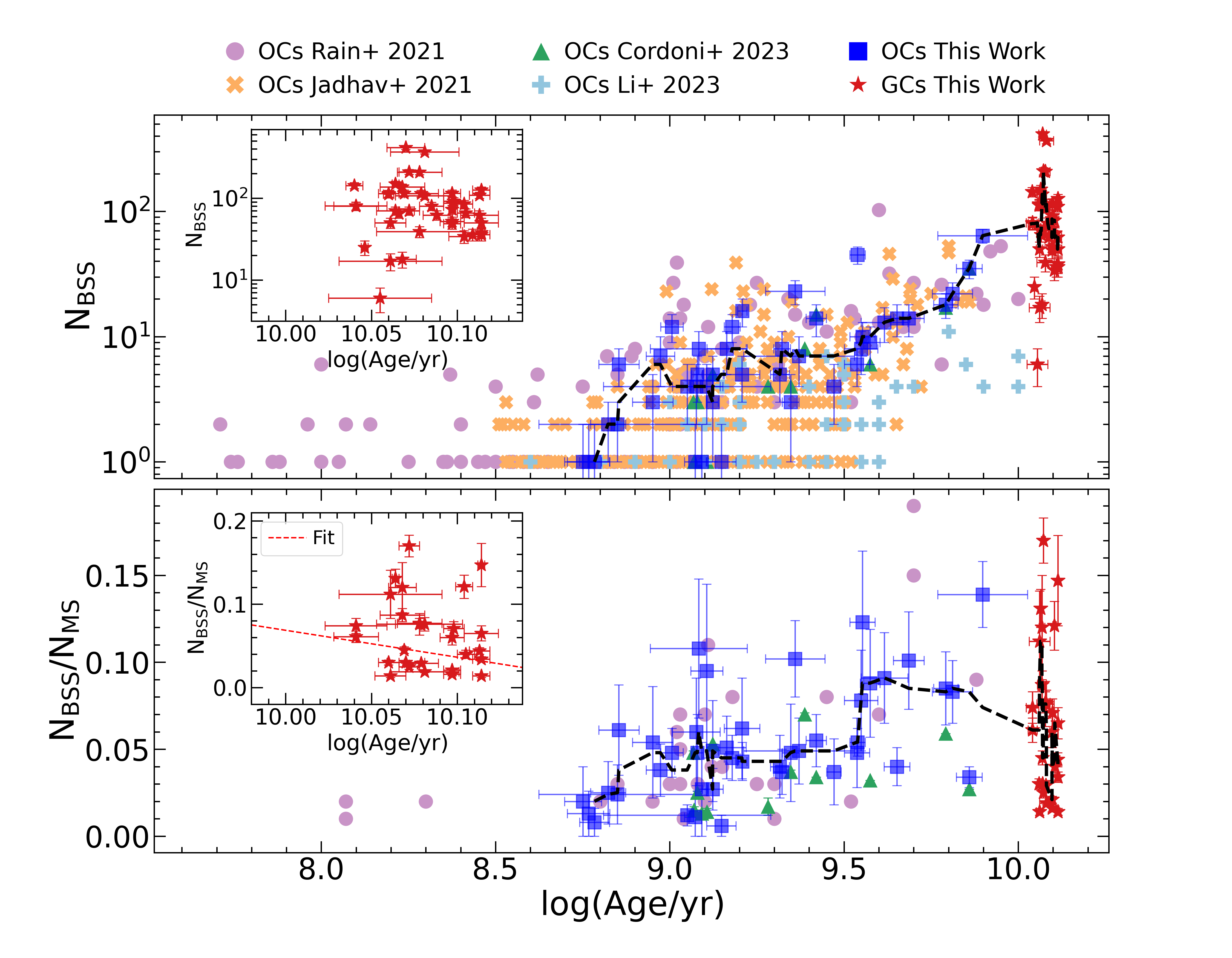}
    \caption{\textit{Top panel:} Absolute number of BSSs vs.~age of parent cluster found in this work and other Gaia-based studies. Uncertainties in the BSS number are assumed as Poisson errors. The inset sub-panel shows a zoom-in on the GC age range for better visualization. The dashed black line denotes the ```ridge-line'' based on the sliding histogram (see text for details). \textit{Bottom panel:} Ratio of BSS numbers relative to the number of MS stars in the luminosity range between MSTO and MSTO+1\,mag (see Sect.~\ref{sec:N_BSS}). We compare the ratio against those studies that also provide this parameter with the same criteria for MS stars number selection, if available. From the considered studies (see legend), only \citetalias{rain_2021} and \citet{cordoni_2023} have this parameter available. Data based in our work is plotted only for those GCs whose ${\rm N}_{\rm MS}$ is not cut by the photometric depth selection described in Section~\ref{subsect:data_sel}; this is done to guarantee a consistent comparison between different clusters and samples.~For the uncertainty in ${\rm N}_{\rm BSS}/{\rm N}_{\rm MS}$ we assume a simple error propagation with Poisson errors for MS stars and BSSs. Errors in age for GCs are based on the difference between our best-fitting isochrone models and those derived from \citet{baumgardt_2023}, whereas errors in ages for OCs are taken from \citet{dias_2021}. Similar to the top figure, the black-dashed line represents the ```ridge-line'' based on the sliding histogram.}
    \label{fig:n_bss_and_fraction}
\end{figure*}

\section{Results and discussion}\label{sec:results}

\subsection{Absolute numbers and fractions of BSSs}\label{sec:N_BSS}

Figure~\ref{fig:n_bss_and_fraction} shows the absolute number (${\rm N}_{\rm BSS}$) of detected BSSs and their number ratio relative to MS stars ($f_{\rm BSS}^{\rm MS}={\rm N}_{\rm BSS}/{\rm N}_{\rm MS}$) in each of our sample GCs and OCs as a function of the parent cluster age.~For the denominator of $f_{\rm BSS}^{\rm MS}$, we use the number of MS stars in the luminosity range between the MSTO and MSTO+1\,mag, similar to previous studies (\citetalias{jadhav_2021}, \citealt{cordoni_2023}).~We select MS stars using the same criteria for all clusters shown in the bottom panel of Figure~\ref{fig:n_bss_and_fraction} to guarantee a homogeneous comparison. Due to the larger average distances of GCs and therefore, their fainter MSTOs, our earlier photometric depth cut (see Sect.~\ref{sec:bss_selection_criteria}) collides with the fainter bound of the $N_{\rm MS}$ luminosity range definition for some distant GCs.~For this reason we inspected by eye every CMD in our GC sample and only provide $f_{\rm BSS}^{\rm MS}$ values for those clusters whose MS stars at $1\,\rm{mag}$ fainter than the MSTO are within the photometric depth bounds. With this limitation, we are able to obtain $f_{\rm BSS}^{\rm MS}$ values for 21 of 41 sample GCs, which is the reason why the number of GCs in the bottom panel of Figure~\ref{fig:n_bss_and_fraction} is smaller than the corresponding number in the upper panel. We also list the values of ${\rm N}_{\rm BSS}$, ${\rm N}_{\rm MS}$ and $f_{\rm BSS}^{\rm MS}$ in \autoref{tab:cluster_param} along with other stellar parameters of the clusters.

The inset panels show zoom-ins at the oldest ages to better visualize any localized trends for the GC subsample.~In \autoref{fig:n_bss_and_fraction}, we also compare the number of BSSs identified in this work against other Gaia-based studies from the literature, i.e.~\citetalias{rain_2021}, \citetalias{jadhav_2021}, \citet{li_2023} and \citet{cordoni_2023}. We point out that our sample is the first that homogeneously studies BSSs in GCs and OCs.~We observe that in all studies, including ours, the BSS population in OCs tends to increase in absolute and fractional numbers towards older cluster ages. This trend appears to reach a maximum at GC ages ($\log[{\rm Age}]\gtrsim 10$), however, and it might even be consistent with the numbers of identified BSSs decreasing with increasing age slightly. 
We also perform sliding histograms of the distributions of (${\rm N}_{\rm BSS}$) and $f_{\rm BSS}^{\rm MS}$ (dashed line) in \autoref{fig:n_bss_and_fraction}. ${\rm N}_{\rm BSS}$ shows a clear trend of increasing with age, 
while the $f_{\rm BSS}^{\rm MS}$ distribution appeared as a plateau in the range $\log[{\rm Age}] \simeq 8.9-9.5$ before suddenly rising again with the varying BSS fraction at older ages. We also perform the Pearson's r-coefficients in the Monte Carlo method \footnote{ Pearson's test measures whether
a pair of two variables are linearly correlated or not. The r-coefficient is defined as the covariance of the variables divided by the product of their standard deviations, and the value ranges from $-$1 to $+$1. $+$1 signifies perfect positive linear correlation between the variables, $-$1 signifies perfect linear anti-correlation, and 0 means no linear correlation. Based on \citet{cohen1988statistical}, coefficient values $>$0.5 ($<-$0.5) indicate a strong positive (negative) linear relation, values between 0.3 and 0.5 ($-$0.3 and $-$0.5) indicate a moderate positive (negative) linear relation, and values between 0 and 0.3 (0 and $-$0.3) indicate a weak positive (negative) linear relation. The Pearson's test also provides a p-value that indicates whether the observed correlation is statistically significant or not. It depends on the number of data points or the sample size. A smaller p-value (generally $<$0.05) suggests that the correlation is statistically significant, while for larger values ($\ge$0.05), we can not conclude a significant correlation \citep{Wasserstein02042016}.} (to include observational errors) for both distributions \citep{Curran_2014_monte_carlo_spearman, Privon_2020_Monte_Carlo_pearson_coeff}\footnote{https://github.com/privong/pymccorrelation?tab=readme-ov-file} and find that the values are 0.67$^{+0.01}_{-0.01}$ for ${\rm N}_{\rm BSS}$ and 0.45$^{+0.02}_{-0.02}$ for $f_{\rm BSS}^{\rm MS}$.\footnote{The numbers in all cases denote the correlation coefficient 50th percentile. The lower and upper errors denote the 16th and 75th percentiles, respectively.} The p-values for both cases are found to be 0. 
The coefficients and p-values suggest that ${\rm N}_{\rm BSS}$ and $f_{\rm BSS}^{\rm MS}$, both correlate with the age of the clusters. Although the correlation for $f_{\rm BSS}^{\rm MS}$ is moderate, one cannot deny the sudden rise of BSS fraction at $\log[{\rm Age}] \sim 9.5$. 
The plateau in $f_{\rm BSS}^{\rm MS}$ in the range $\log[{\rm Age}] \simeq 8.9-9.5$ could be because, at a young age, massive stars reach larger radii in the AGB phase and mass transfer is possible even in wide binaries, which favours BSS formation to make a constant BSS fraction with respect to the MS population \citep[][]{leiner_2021}. Other possible mechanisms of more BSS formation are wind accretion from the AGB companion and stable mass transfer in the subgiant branch or Hertzprung gap in younger clusters.  
The increase in the BSS fractions at older ages might be due to the longer MS lifetime of low-mass BSSs in older clusters and/or the contribution from collisional BSSs, however \citep[see their Fig.5]{leiner_2021}.

In the inset, we highlight the distribution of $f_{\rm BSS}^{\rm MS}$ for GCs and compare the trend with that of OCs. 
We notice that for OCs, as they get older, the $f_{\rm BSS}^{\rm MS}$ values increase, whereas for GCs, as they become older, this fraction tends to decrease. We also added a linear fit to the inset plot as a dashed red line that supports this observation. We find the best fit to be:
\begin{equation}
    f_{\rm BSS, GC}^{\rm MS} = (-0.364 \pm 0.089) \times \log({\rm Age}/{\rm yr}) + (3.731 \pm 0.902).
\end{equation} 
Pearson's r-coefficient ($-0.17^{+0.01}_{-0.01}$) suggests, however, that the anti-correlation is rather weak. A similar weak anti-correlation ($-0.12^{+0.01}_{-0.0}$) is also found for ${\rm N}_{\rm BSS}$ as well. The p-values are found to be 0 for both cases. As the very central regions of the GCs are excluded in this study because of Gaia's limitation, incorporating HST observations alongside Gaia data would be beneficial for investigating this correlation. At the same time, we need to include more number of GCs with deep photometry, as we missed to include some GCs due to the magnitude limitation of Gaia.

We note that the dynamic range in $f_{\rm BSS}^{\rm MS}$ covered by OCs and GCs is similar, unlike the absolute number of BSSs, where GCs show larger absolute numbers due to their larger masses, i.e.~more numerous stars. More specifically, GCs show $\langle f^{\rm MS}_{\rm BSS} \rangle_{\rm GC} = 0.07 \pm 0.04$ while for OCs we find a slightly larger mean ratio of $\langle f^{\rm MS}_{\rm BSS} \rangle_{\rm OC} = 0.08 \pm 0.05$.

We do not detect BSSs in clusters younger than $\sim\!500\,{\rm Myr}$ in our sample, supporting the results found by \citetalias{rain_2021}, \citetalias{jadhav_2021} and \citet{cordoni_2023}.

\subsection{Astrophysical parameters of BSSs} \label{sec:bss_astrophysical_params}

We derive the following astrophysical parameters for our sample BSSs: effective temperature ($T_{\rm eff}$), stellar mass ($M$), luminosity ($L$) and surface gravity ($\log g$). 
We used three different approaches: the infrared flux method (IRFM; from color-temperature relations), the isochrone best--fit method (from PARSEC isochrones) and the XPP from Gaia DR3 spectra, if available. While the isochrone method provides all astrophysical parameters, the IRFM method provides $T_{\rm eff}$ and the XPP method gives $T_{\rm eff}$ \& $\log g$. We mainly use the parameters obtained from the isochrone method, but it is necessary to compare $T_{\rm eff}$ values, obtained from all three methods to check the consistency in the parameter estimations. In~\autoref{sec:analysis}, we describe all three methods and the comparison of the $T_{\rm eff}$ values obtained from them, in detail. In the following subsections, we summarize the results and compare the stellar parameters with the literature values.

\subsubsection{Effective temperature}

We notice that $T_{\rm eff}$ values extracted from different approaches range from $\sim$ 6000 K to 9000 K and appear consistent with each other with a little systematic offset towards higher temperature ($T_{\rm eff} >$ $\sim\!8000\,{\rm K}$).
From these comparisons, we estimate that the systematic uncertainty between various methods is of the order $\Delta T_{\rm eff}\simeq100$\,K. 
From the IRFM we find $\langle {\rm T}_{\rm eff, IRFM}\rangle = (6772 \pm 630)~{\rm K}$ for GCs and $\langle {\rm T}_{\rm eff, IRFM}\rangle = (7598 \pm 1476)~{\rm K}$ for OCs. From isochrone fitting we find $\langle {\rm T}_{\rm eff, isochrone}\rangle = (6829 \pm 529)~{\rm K}$ for GCs and $\langle {\rm T}_{\rm eff, isochrone} \rangle = (7549 \pm 1331)~{\rm K}$ for OCs. Only 537 of 4399 BSSs ($\sim\!12\%$ of our sample) have XPP parameters measured for 279 stars in GCs and 258 stars in OCs. With this method, we find the mean BSS effective temperature of $\langle {\rm T}_{\rm eff, XPP}\rangle = (7271 \pm 732)~{\rm K}$ for GCs and $\langle {\rm T}_{\rm eff, XPP} \rangle = (7269 \pm 1331)~{\rm K}$ for OCs. 
We list these parameters in \autoref{tab:bss}, and the full table is available at the CDS.

\autoref{table:comparison} compare our derived values for $T_{\rm eff}$ with those found in the literature for various OCs and GC, as well as BSS samples. We find that our effective temperatures derived are in very good agreement with previous studies. For example, \citet{jadhav_2019}, \citet{sindhu_2019} and \citet{pandey_2021} have analyzed different BSSs in the OC NGC\,2682 (also known as M67) by fitting models to spectral energy distributions (SEDs) searching for BSS hot companions; these studies derive BSS temperatures in the range between $6510$ and $8300~{\rm K}$, which are in excellent agreement with our measurements, yielding a mean $T_{\rm eff} = 7300 \pm 1000~{\rm K}$ for all methods combined. We also point out the excellent agreement with the $T_{\rm eff}$ values derived in \citet{gosnell_2015}, where the authors find $\langle T_{\rm eff} \rangle = 6396 \pm 23\,{\rm K}$ for 19 BSSs, which is in very good agreement with our results independent of the applied method ($6366 \pm 49\,{\rm K}$ and $6473 \pm 41\,{\rm K}$, for the IRFM and isochrone method, respectively). Finally, our $\langle T_{\rm eff} \rangle \approx 6800~{\rm K}$ is also in very good agreement with theoretical predictions for BSSs in GCs from \citet{stepien_2015} who calculate $\langle T_{\rm eff} \rangle = (6969 \pm 96)\,{\rm K}$, and who had applied different initial conditions to binary systems.

\subsubsection{Stellar mass and age of the BSSs} \label{mass_age}
Stellar mass and age are important parameters in the study of the evolutionary state of BSSs. We derive these parameters along with equivalent stellar luminosity and stellar surface gravity for all BSSs in our sample based on the isochrone fitting method. We obtain the mean luminosity $\langle \log (L_{\rm BSS}/L_\odot) \rangle = 0.58 \pm 0.21$ for GCs and $\langle \log (L_{\rm BSS}/L_\odot) \rangle = 1.13 \pm 0.44$ for OCs, and find the mean surface gravity of $\langle \log (g_{\rm BSS}/[{\rm cm} \cdot {\rm s}^{-2}]) = 4.14 \pm 0.22$ for GCs and $\langle \log (g_{\rm BSS}/[{\rm cm} \cdot {\rm s}^{-2}]) = 3.98 \pm 0.21$ for OCs.~These values correspond to a mean stellar mass $\langle M_{\rm BSS} \rangle = 1.02 \pm 0.1~M_\odot$ for BSSs in GCs and $\langle M_{\rm BSS} \rangle = 1.75 \pm 0.45~M_\odot$ for BSSs in OCs, which altogether translates to a mean stellar age of $\log ({\rm Age}_{\rm BSS}/{\rm yr}) = 9.586 \pm 0.318$ for BSSs in GCs and $\log ({\rm Age}_{\rm BSS}/{\rm yr}) = 9.065 \pm 0.344$ for BSSs in OCs. 

We compare these values with previous results from the literature in \autoref{table:comparison}. We find that the average mass of our BSS sample in GCs ($\langle M_{\rm BSS, GC} \rangle = 1.02 \pm 0.1 \ M_\odot$) differs from previous works, which report $\langle M_{\rm BSS} \rangle = 1.7\pm 0.4 \ M_\odot$ \citep{shara_1997} %\tpc{<-- Not in table!} 
and $\langle M_{\rm BSS} \rangle = 1.22 \pm 0.12 \ M_\odot$ \citep{baldwin_2016}. This may be partly explained by the lower BSS numbers of previous studies, but also by either of the previous arguments, i.e.~i) limitations of the Gaia data selection and ii) BSS mass segregation in GCs. In particular, the latter argument lends credence to the reason for this offset as the two previous studies are based on HST observations that target the GC centers. Our sample, on the other hand, is based on BSS samples from outside the half-light radius. Nevertheless, as can be appreciated in Table~\ref{table:comparison}, our BSS mass results are in very good agreement with those measured by \citet[][$\langle M_{\rm BSS}\rangle = 1.07 \pm 0.13\,M_\odot$]{simunovic_2017}, \citet[][$\langle M_{\rm BSS} \rangle = 1.06 \pm 0.09\,M_\odot$]{fiorentino_2014} and \citet[][$\langle M_{\rm BSS}\!\approx\!1\!-\!1.1\,M_\odot$]{stepien_2015}. The comparison with \cite{simunovic_2017} is in particular striking, as this work measured the parameters for 1507 BSSs in 38 GCs using the isochrone fitting technique similar to our procedure, which makes it statistically the most robust comparison.

\begin{table*}
  \centering
  \caption{Comparison between BSS astrophysical parameters obtained in this study against those found in the literature. }\label{table:comparison}
  \begin{tabular}{@{} *7c @{}}
    \toprule
    & Name & ${\rm N}_{\rm BSS}$ & $\langle T_{\rm eff}\rangle$ & $\langle \log(g) \rangle$ & $\langle {\rm Mass} \rangle$ & Reference\\ 
    &  & & $({\rm K})$ & & $(M_\odot)$ & \\\midrule
    Open Clusters & & & & & &\\\midrule
    & NGC 188  & 19  & $6396 \pm 233$ & \textemdash  & \textemdash & \citet{gosnell_2015}\\
    &  & 21  & $6366 \pm 496$ & \textemdash  & \textemdash & This Work -- IRFM\\
    &  & 21  & $6473 \pm 416$ & $4.05 \pm 0.17$  & $1.45 \pm 0.17$ & This Work -- isochrone\\
    & NGC 2682  & 21  & $6510 \pm 429$ & \textemdash & \textemdash & \citet{jadhav_2019}\\
    &  & 3  & $8333 \pm 721$ & \textemdash & \textemdash & \citet{sindhu_2019}\\
    &  & 6  & $7291 \pm 485$ & $3.67 \pm 0.52$  & \textemdash & \citet{pandey_2021}\\
    &  & 9  & $7308 \pm 1220$ & \textemdash & \textemdash & This Work -- IRFM\\
    &  & 9  & $7287 \pm 1015$ & $4.08 \pm 0.20$ & $1.70 \pm 0.39$ & This Work -- isochrone\\
    & NGC 7789  & 15  & $8433 \pm 961$  & $4.23 \pm 0.46$ & \textemdash & \citet{vaidya_2022}\\
    &  & 16  & $8690 \pm 1142$  & \textemdash & \textemdash & This Work -- IRFM\\
    &  & 16  & $8469 \pm 1008$  & $3.93 \pm 0.22$ & $2.03 \pm 0.31$ & This Work -- isochrone\\
    & Multiple OCs  & 1368  & \textemdash  & \textemdash & $2.49 \pm 1.13$ & \citet{jadhav_2021}\\
    &  & 434  & $7598 \pm 1476$  & \textemdash & \textemdash & This Work -- IRFM\\
    &  & 434  & $7549 \pm 1331$  & $3.98 \pm 0.21$ & $1.75 \pm 0.45$ & This Work -- isochrone\\\midrule
    Globular Clusters & & & & & & \\\midrule
    & NGC 362  & 26  & $7810 \pm 863$  & $4.29 \pm 0.53$ & \textemdash & \citet{dattatrey_2023}\\
    &  & 137  & $6718 \pm 465$  & \textemdash & \textemdash & This Work -- IRFM\\
    &  & 137  & $6816 \pm 405$  & $4.09 \pm 0.20$ & $0.94 \pm 0.08$ & This Work -- isochrone\\
    & NGC 5272  & 7  & $7093 \pm 917$  & $3.61 \pm 0.58$ & \textemdash & \citet{de_marco_2005}\\
    &  & 210  & $6910 \pm 621$  & \textemdash & \textemdash & This Work -- IRFM\\
    &  & 210  & $6978 \pm 520$  & $4.11 \pm 0.22$ & $0.98 \pm 0.10$ & This Work -- isochrone\\
    & NGC 6541  & 9  & \textemdash  & \textemdash & $1.06 \pm 0.09$ & \citet{fiorentino_2014}\\
    &  & 52  & $7240 \pm 760$  & $4.09 \pm 0.26$ & $1.00 \pm 0.10$ & This Work -- isochrone\\
    & Multiple GCs  & 64  & $6969 \pm 967$  & \textemdash & $1.11 \pm 0.09$ & \citet{stepien_2015}\\
    &   & 598  & \textemdash  & \textemdash & $1.22 \pm 0.12$ & \citet{baldwin_2016}\\
    &   & 1507  & \textemdash  & \textemdash & $1.07 \pm 0.13$ & \citet{simunovic_2017}\\
    & & 3965 & $6772 \pm 630$ & \textemdash & \textemdash & This Work -- IRFM\\
    & & 3965 & $6829 \pm 529$ & $ 4.14 \pm 0.22$ & $1.02 \pm 0.01$ & This Work -- isochrone\\\bottomrule
  \end{tabular}
  
  \vspace{2mm}
  
  {\raggedright\textbf{Notes.} Columns are as follows: 1) Cluster name. 2) Number of available BSSs in the literature with available astrophysical parameters. 3) Mean of the effective BSS temperature provided. 4) Mean of the decadal logarithm of the BSS surface gravity. 5) Mean of the BSS stellar mass. 6) Reference. For data provided by \citet[][or \citetalias{jadhav_2021}]{jadhav_2021}, we considered stars labeled BSS and pBSS (probable BSS).
  \par}
\end{table*}

The average BSS mass for our OC sample ($\langle M_{\rm BSS, OC} \rangle = 1.75 \pm 0.45\,M_\odot$ compares favorably with the corresponding value found by \citetalias{jadhav_2021} considering the uncertainties. They report an average mass value of $\langle M_{\rm BSS, OC}\rangle = 2.67 \pm 1.2 \ M_\odot$ if only confirmed BSSs from their catalog are considered, and $\langle M_{\rm BSS, OC}\rangle = 2.49 \pm 1.13\,M_\odot$ if both confirmed and candidate/probable BSSs listed in their catalog are included. As was explained previously, the difference between these works is likely the result of our Gaia ${\rm G}$ magnitude cut-off criterion, discarding all bright stars with $G_{\rm BP} \leq 10$ as they are likely saturated and therefore, their Gaia parameters likely affected by systematic errors. This not only explains the difference between our results, as well as why our dispersion in BSS mass for OCs is smaller, but also the lack of BSSs younger than $\sim500$ Myr in our sample, most of which would naturally be among the most massive BSSs.

\subsection{ Effective temperature versus cluster age}

\begin{figure}
\includegraphics[width=\columnwidth]{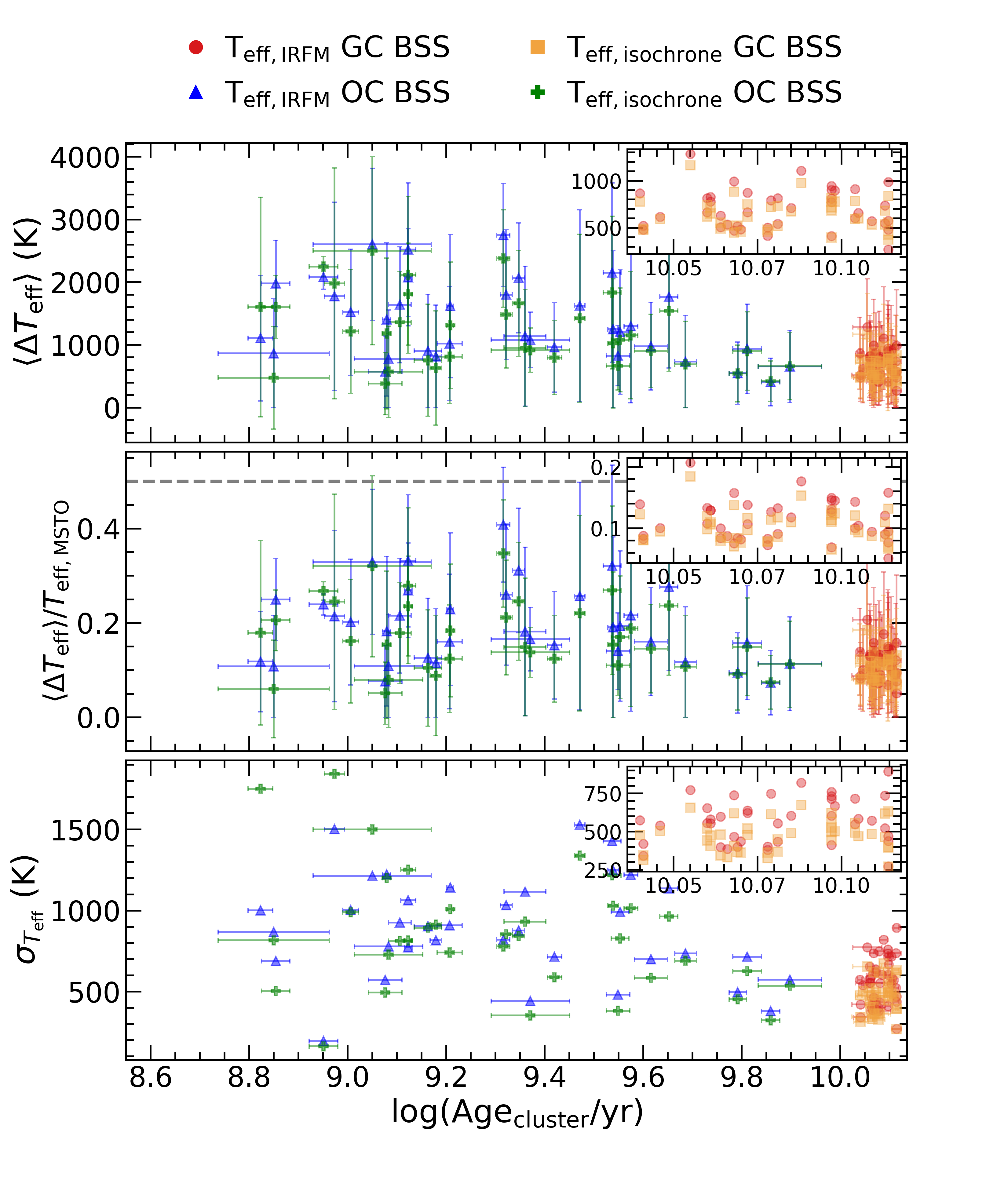}
    \caption{$T_{\rm eff}$ vs. cluster age. \textit{Top:} $\langle T_{\rm eff} \rangle$ –-the mean of the difference between BSS $T_{\rm eff}$ and MSTO $T_{\rm eff}$, found with the same method, for stars within the same cluster–- vs. cluster age. Blue triangles and red dots represent $T_{\rm eff}$ obtained from color–temperature relations for OCs and GCs, respectively. Green plus symbols  (\texttt{+}) and orange squares represent $T_{\rm eff}$ derived from isochrone models for OCs and GCs, respectively. 
    At the top right corner we show the results only for GCs without error bars for a better appreciation, keeping the same axes from the major plot. Error bars along vertical axis show the dispersion of BSSs $T_{\rm eff}$. 
    \textit{Middle:}  difference between $\langle T_{\rm eff} \rangle$ and the MSTO $T_{\rm eff}$, normalized by the MSTO temperature. Horizontal gray dashed line indicates ${\rm BSS} \ T_{\rm eff} = 1.5 \times {\rm MSTO} \ T_{\rm eff}$. \textit{Bottom:} Dispersion of BSS $T_{\rm eff}$ vs. cluster age.}
    \label{fig:age_cluster_vs_teff}
\end{figure}

In Figure~\ref{fig:age_cluster_vs_teff} we compare the average MSTO offset in $T_{\rm eff}$ of BSS in a cluster against the corresponding cluster age. To compare self-consistently all the BSSs with respect to the same evolutionary reference point, given the clusters' very different ages and metallicities, we use the $T_{\rm eff}$ of their MSTO as a reference. For each BSS, we obtain $\Delta T_{\rm eff} = T_{\rm eff, BSS} - T_{\rm eff, MSTO}$. We then compute $\langle \Delta T_{\rm eff} \rangle$ which is the mean difference between the BSS temperatures and the $T_{\rm eff, MSTO}$ in each cluster. As XPP parameters are not available for stars at the MSTO, we only do this for $T_{\rm eff}$ obtained by isochrones and color--temperature relation, i.e.~IRFM, and recall the systematic offsets found in the comparison of these methods (see Fig.~\ref{fig:effective_temperature_relations_methods}).~The error bars in $\Delta T_{\rm eff}$ represent the dispersion of BSS temperatures in each cluster, while the error bars on the horizontal axis are age uncertainties.

In the top panel of figure~\ref{fig:age_cluster_vs_teff}, BSSs in OCs show on average a trend against the parent cluster age where younger OCs show larger $\langle \Delta T_{\rm eff} \rangle$ offsets, i.e.~hotter BSSs, with respect to those found in older OCs, or in other words, the difference between the effective temperature of BSSs and the MSTO effective temperature decreases with increasing parent cluster age. 
Pearson's coefficients for IRFM (isochrone) method is $-0.2_{-0.13}^{+0.13}$ ($-0.2_{-0.14}^{+0.14}$), suggesting the correlation to be weak, which could be due to a large dispersion in the BSS temperatures in the younger cluster. 
We expect the observed trend, however, because younger clusters host more massive stars which are also larger in diameter \citep[e.g.][]{Iben13-2}, and therefore, have likely more efficient MT between binary members and higher collision/capture probabilities with other stars. These hotter BSS stars evolve faster than BSS stars around the MSTO and, thus, the difference in average $T_{\rm eff}$ between BSSs and BSSs around MSTO is expected to decrease with time. We also notice that the dispersion in $\Delta T_{\rm eff}$ decreases with the age of the OC, which may indicate that BSSs tend to be more concentrated in the color-luminosity space as the OC ages, which is also shown in the bottom panel of the figure. Pearson's coefficient values further confirm the trend: $-0.16_{-0.02}^{+0.02}$ and $-0.35_{-0.03}^{+0.02}$ for IRFM and isochrone methods, respectively. 

We do not observe any of these correlations for BSSs in GCs, which is also supported by the Pearson's coefficient for the IRFM (isochrone) method $-0.01_{-0.15}^{+0.14}$ ($-0.02_{-0.15}^{+0.15}$) (see inset panels in Fig.~\ref{fig:age_cluster_vs_teff}). We recall, however, that we might be exposed to biases for the GC subsample: i) we applied a color cut when selecting BSSs in GCs to avoid including BHB stars (see Fig.~\ref{fig:bss_selection_example} and Sect.~\ref{sec:bss_selection_criteria}) which, although unlikely may cut into the distribution of BSSs with extremely large $\Delta T_{\rm eff}$ (this could in principle be tested with UV photometry); ii) Gaia is only capable to observe stars outside $\sim\!r_{\rm hl}$ (half-light radius) in GCs, which could change the mean $\Delta T_{\rm eff}$ if we assume that mass segregation has already acted on the BSS population \citep{contreras_ramos_2012, ferraro_2012}.~Without consideration of these potential biases and taking the distributions in Figure~\ref{fig:age_cluster_vs_teff} at face value, this means that as GCs age past $\sim\!10$\,Gyr, BSSs are more or less equally distributed in $\Delta T_{\rm eff}$.

In the middle panel of figure~\ref{fig:age_cluster_vs_teff}, we check if the fractional $\langle \Delta T_{\rm eff} \rangle$ excess with respect to $T_{\rm eff, MSTO}$ has any correlation with the cluster age.  
We do not find any trend for either OCs or GCs, which is also supported by the Pearson's coefficient test. Pearson's coefficients for IRFM (isochrone) are $-0.09_{-0.13}^{+0.13}$ ($-0.1_{-0.14}^{+0.14}$) for OCs and $-0.02_{-0.14}^{+0.14}$ ($-0.01_{-0.16}^{+0.15}$).

\begin{figure*}
    \includegraphics[width=\textwidth]{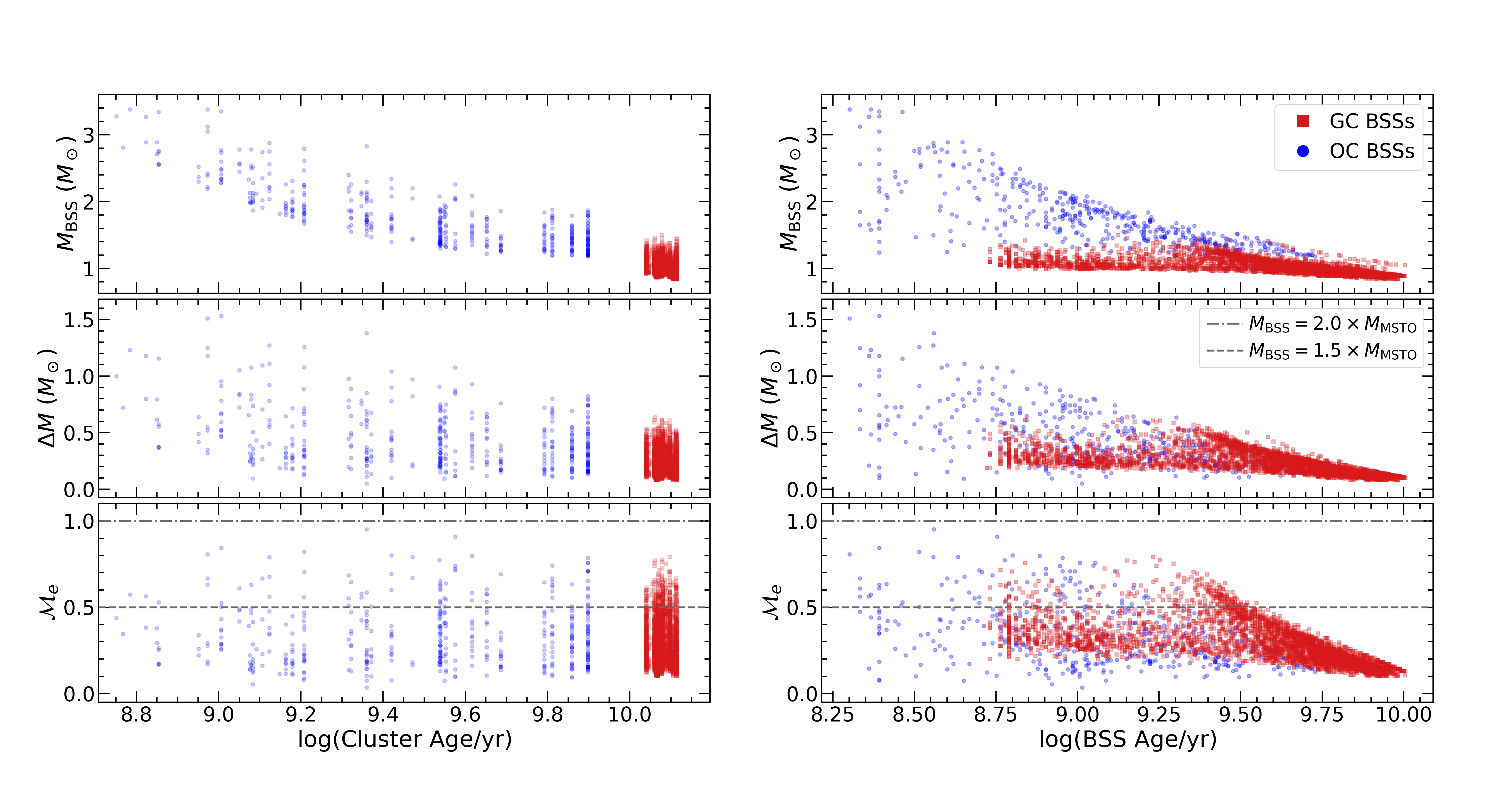}  
    \caption{Mass of BSSs. \textit{Left:} Mass of BSSs (found with isochrone models) vs. parent cluster age. Upper panel shows the mass of the BSSs, the middle panel shows the difference between the BSS mass and the MSTO mass, and the lower panel shows the mass excess factor, $\mathcal{M}_{\rm e}$ --defined by \citetalias{jadhav_2021} as the difference between the BSS Mass and the MSTO Mass, normalized by the MSTO mass--, with respect to the parent cluster age. GC BSSs and OC BSSs are represented by red and blue points, respectively. 
    We additionally draw two horizontal gray lines at the lower panel; dot-dashed gray horizontal line denotes the limit for $M_{\rm BSS} = 2.0 \times M_{\rm MSTO}$, whereas dashed gray horizontal lines denote $M_{\rm BSS} = 1.5 \times M_{\rm MSTO}$. Focusing on the lower panel and following \citetalias{jadhav_2021} classification, all stars between both horizontal lines are classified as high--$\mathcal{M}_{\rm e}$ (most \emph{likely} formed via collisions/mergers), and all stars under the horizontal dashed gray line are classified as low--$\mathcal{M}_{\rm e}$ (most \emph{likely} formed via MT). \textit{Right:} Same as the left side of the figure, but displaying Mass of BSSs vs. BSSs age (found with isochrone models).}
    \label{fig:bss_vs_mass}
\end{figure*}

\subsection{Fractional mass excess of BSSs}

BSSs are expected to gain mass during their MS lifetime. This means that the ``excess'' between their current mass and the stellar mass at the MSTO could provide hints about their formation pathways. Recall that, based on previous observations \citep{dalessandro_2013, dattatrey_2023}, hotter (bluer), but relatively faint BSSs are likely to be formed through collisions, whereas slightly cooler (redder), but slightly brighter BSSs are likely to be formed by MT. Even though this is almost certainly not strictly followed by the universe (modulo stellar-mass loss rate, binary parameters, environmental factors, such as tidal shocks and disruption of binaries, etc.), it could be considered a first scenario for the formation of particular BSS subpopulations. Trying to provide a quantity that could give a hint which pathway dominates certain evolutionary phases, \citetalias{jadhav_2021} defined a parameter called the ``fractional mass excess'', or $\mathcal{M}_{\rm e}$. This quantity is defined as
\begin{equation}\label{eq:fractional_mass_excess}
    \mathcal{M}_{\rm e} = \frac{M_{\rm BSS} - M_{\rm MSTO}}{M_{\rm MSTO}},
\end{equation}
where $M_{\rm BSS}$ is the mass of the BSS and $M_{\rm MSTO}$ is the stellar mass at the MSTO, which in our case is obtained from isochrone models. In simple words, $\mathcal{M}_{\rm e}$ is the excess between the BSS mass and MSTO mass, normalized by the MSTO mass. Note that $\mathcal{M}_{\rm e} = 0.5$ is equivalent to $M_{\rm BSS} = 1.5 \times M_{\rm MSTO}$, and $\mathcal{M}_{\rm e} = 1.0$ corresponds to $M_{\rm BSS} = 2 \times M_{\rm MSTO}$. By definition, this ``fractional mass excess'' is roughly equivalent to the MT efficiency if we assume that both member stars of the system, the donor and the secondary star, are MS stars \citep{shao_2016}. \citetalias{jadhav_2021} decided to divide the $\mathcal{M}_{\rm e}$ range into three ad-hoc regimes: low--$\mathcal{M}_{\rm e}$ ($\mathcal{M}_{\rm e} < 0.5$), high--$\mathcal{M}_{\rm e}$ ($0.5 < \mathcal{M}_{\rm e} < 1.0$) and extreme--$\mathcal{M}_{\rm e}$ ($\mathcal{M}_{\rm e} > 1.0$) where, as a first approximation they speculated, different BSS formation pathways would dominate, which are MT, stellar collisions, and multiple-mergers/MT, respectively.

To make headway in the discussion of BSS formation pathways, we study the behavior of $\mathcal{M}_{\rm e}$ as a function of cluster age and the equivalent BSS age from isochrone fitting, both of which are displayed in Figure~\ref{fig:bss_vs_mass} along with the BSS mass and mass offset scalings. We focus first on the left-side panel. In the top panel, we observe that, as expected, younger clusters host BSSs with higher masses, which decrease with cluster age. The Pearson coefficient also provides a strong correlation value of $-0.88_{-0.0}^{+0.0}$, combining both OCs and GCs. This correlation is consistent with the decreasing MSTO mass as a stellar population ages, since the lifetime of a star, $\tau$, scales as $\tau \propto \frac{M}{L} \simeq \frac{1}{M^3}$, ~i.e.~more massive stars have $M^{-3}$ shorter lifetimes compared to less massive stars, which translates into the BSS mass relation in the top left panel of Figure~\ref{fig:bss_vs_mass}. This correlation also appears stronger when only OCs are considered ($-0.8_{-0.0}^{+0.01}$), but this is not true for the GC sample ($-0.1_{-0.01}^{+0.01}$). The reasons for GCs are the same as discussed in the previous sections.   

The middle panel on the left side of Figure~\ref{fig:bss_vs_mass} shows the difference between the BSS mass and the MSTO mass as a function of cluster age. We see that the BSS with higher mass difference are absent towards the older cluster age with Pearson coefficient of $-0.5_{-0.01}^{+0.0}$. This is also expected from regular stellar evolution since more massive stars end their life disproportionately earlier while the MSTO mass decreases more slowly with increasing age, which is exactly what is observed, i.e. $\Delta M = M_{\rm BSS}-M_{\rm MSTO}$ shrinks as time advances. We also notice here that GC sample alone does not show any correlation ($-0.02_{-0.01}^{+0.01}$).

More surprisingly, we find no trend of $\mathcal{M}_{\rm e}$ with cluster-age (bottom left panel in Fig.~\ref{fig:bss_vs_mass}) either for OCs or GCs with Pearson coefficient values $0.02_{-0.02}^{+0.02}$ and $0.01_{-0.01}^{+0.01}$, respectively. We observe, however, that the fraction of high-$\mathcal{M}_{\rm e}$ ($0.5 < \mathcal{M}_{\rm e} < 1.0$) BSSs is larger in OCs ($18.7\%$) compared to GCs ($5.8\%$), which may indicate that the collisional formation channel is more pronounced in younger star clusters. This in turn might be related to sufficiently old dynamical ages of star cluster, which can, in principle, have undergone one or more core collapse, boosting the stellar encounter rates, and therefore, the formation of collision-induced BSS formation. This might be accomplished for star clusters with ages of a few billion years, however, and it requires further investigation and modeling.
We find $81.3\%$ and $18.7\%$ as the BSS fractions with low and high--$\mathcal{M}_{\rm e}$ (split at $\mathcal{M}_{\rm e}=0.5$), respectively, for OCs. In GCs, we find these proportions to be $94.3\%$ and $5.8\%$ for low and high--$\mathcal{M}_{\rm e}$, respectively. Considering all clusters, the percentage of low--$\mathcal{M}_{\rm e}$ is $93\%$ and for high--$\mathcal{M}_{\rm e}$ we obtain $7\%$.

Additionally, we plot the BSS mass parameters versus~BSS ages in the right panels of Figure~\ref{fig:bss_vs_mass}. The relations for $M_{\rm BSS}$ and $\Delta M$ are consistent with those found against the cluster age (left side of the same Figure), i.e., both of these parameters decrease as a function of BSS age and stellar population age. In addition, we see various sequences that indicate that the BSSs in OCs and GCs predominantly populate different parameter spaces. A striking distribution is found in the $\mathcal{M}_{\rm e}$ against $\log(\rm {BSSs \ age})$ plot (bottom right panel) which shows two dominant sequences, both of which are consistent with decreasing $\mathcal{M}_{\rm e}$ as BSSs become older, albeit at different rates. We will later show that these sequences are likely representative of the different BSS formation channels. The pearson coefficients for all these three relations for OCs (GCs) are: $-0.75_{-0.0}^{+0.01}$ ($-0.61_{-0.0}^{+0.0}$) for $M_{\rm BSS}$ versus BSS ages; $-0.6_{-0.01}^{+0.01}$ ($-0.62_{-0.01}^{+0.0}$) for $\Delta M$ versus BSS ages; and $-0.41_{-0.01}^{+0.01}$ ($-0.62_{-0.0}^{+0.01}$) for $\mathcal{M}_{\rm e}$ versus BSS ages.

\begin{figure}[!ht]
    \includegraphics[width=\columnwidth]{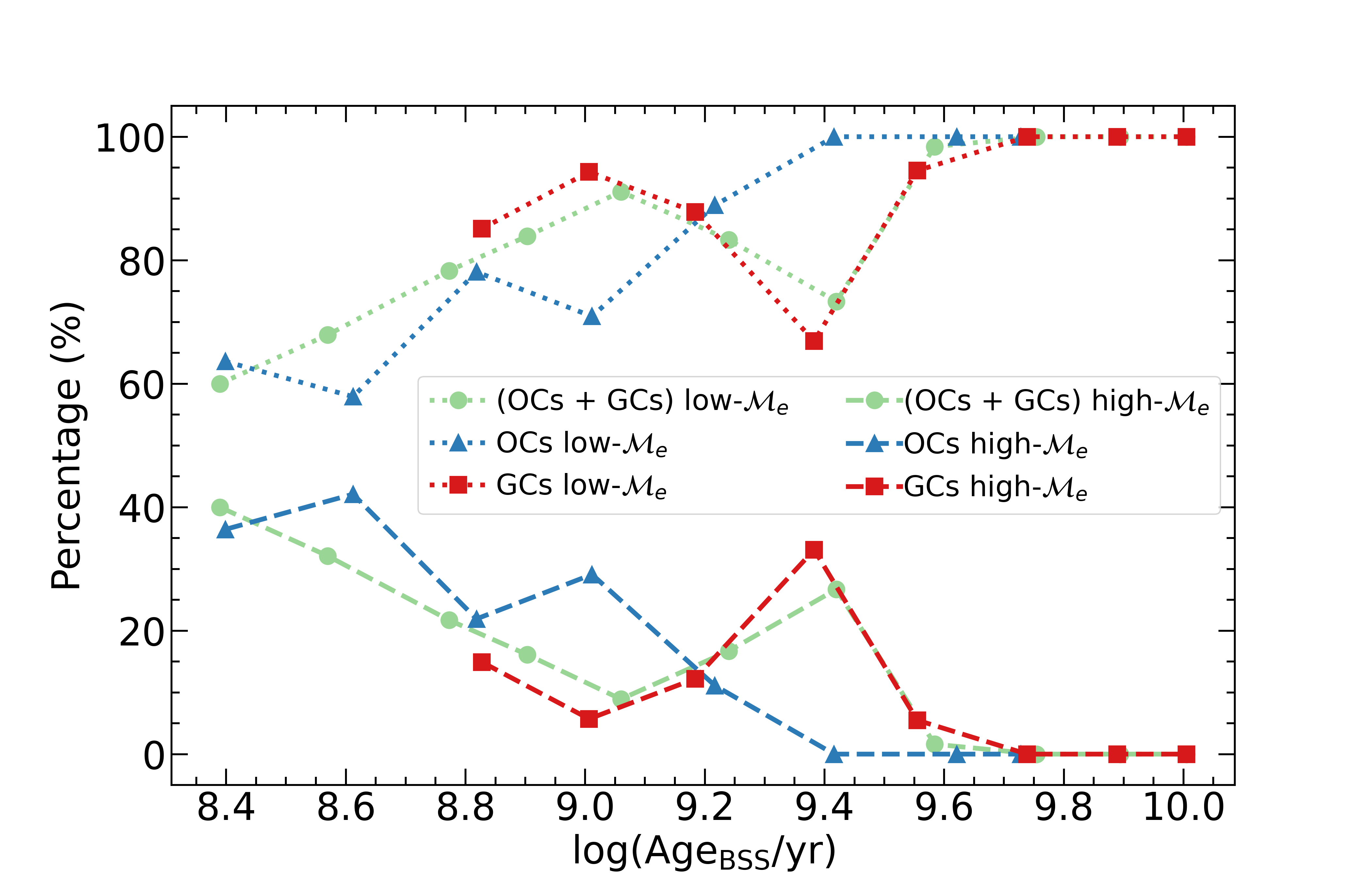}
    \caption{$\mathcal{M}_{\rm e}$ percentage vs. BSSs age (found from isochrone models). All (OCs + GCs) BSSs are symbolized as green circles, GC BSSs are represented as red squares and OC BSSs as blue triangles. The line connecting the symbols indicates the $\mathcal{M}_{\rm e}$ regime: dashed lines indicates high--$\mathcal{M}_{\rm e}$ and dotted lines indicate low--$\mathcal{M}_{\rm e}$.}
    \label{fig:frac_mass_excess_percentage_vs_BSS_age}
\end{figure}

Figure \ref{fig:frac_mass_excess_percentage_vs_BSS_age} shows the percentage of low--$\mathcal{M}_{\rm e}$ and high--$\mathcal{M}_{\rm e}$ BSSs as a function of BSS age. For OCs, it is clear that the high--$\mathcal{M}_{\rm e}$ fraction vanishes above the BSS ages of $\sim\!2-3\,{\rm Gyr}$, whereas the low--$\mathcal{M}_{\rm e}$ fraction dominates, with a percentage $>90\%$ in the same older age range. Curiously, the high--$\mathcal{M}_{\rm e}$ fraction in GCs increases to $\sim 30\%$ for BSS ages between $\sim 1.5$--$2.5 \ {\rm Gyr}$, while the low--$\mathcal{M}_{\rm e}$ fraction experiences a corresponding dip. This may be due to the predominance of BSS formed through the collisional channel at those ages. \cite{ferraro_2009, dalessandro_2013, simunovic_2014} fitted collisional isochrones \citep{sills_2009} to the blue sequence of BSSs, finding that their ages are $\sim\!2 \ {\rm Gyr}$, concluding that double sequences of BSSs in GCs may be products of violent interaction episodes in the past core-collapse events, which is in agreement with our age range found for BSS in GCs where the high--$\mathcal{M}_{\rm e}$ fraction increases and the low--$\mathcal{M}_{\rm e}$ fraction dips. For GCs older than $\sim\!3.5 \ {\rm Gyr}$ the high--$\mathcal{M}_{\rm e}$ fraction rapidly vanishes similar to those for OCs. Overall, we observe that for all BSSs in our sample low/high--$\mathcal{M}_{\rm e}$ total fraction is dominated by OCs for younger BSSs ages ($\sim\!250\!-\!600\,{\rm Myr}$) and by GCs for older BSSs ages ($\sim\!1.5\!-\!10\,{\rm Gyr}$) as expected by their sample characteristics (see Sect.~\ref{sec:N_BSS}).

\subsection{Comparison with \citetalias{jadhav_2021}}
\begin{figure}
    \includegraphics[width=\columnwidth]{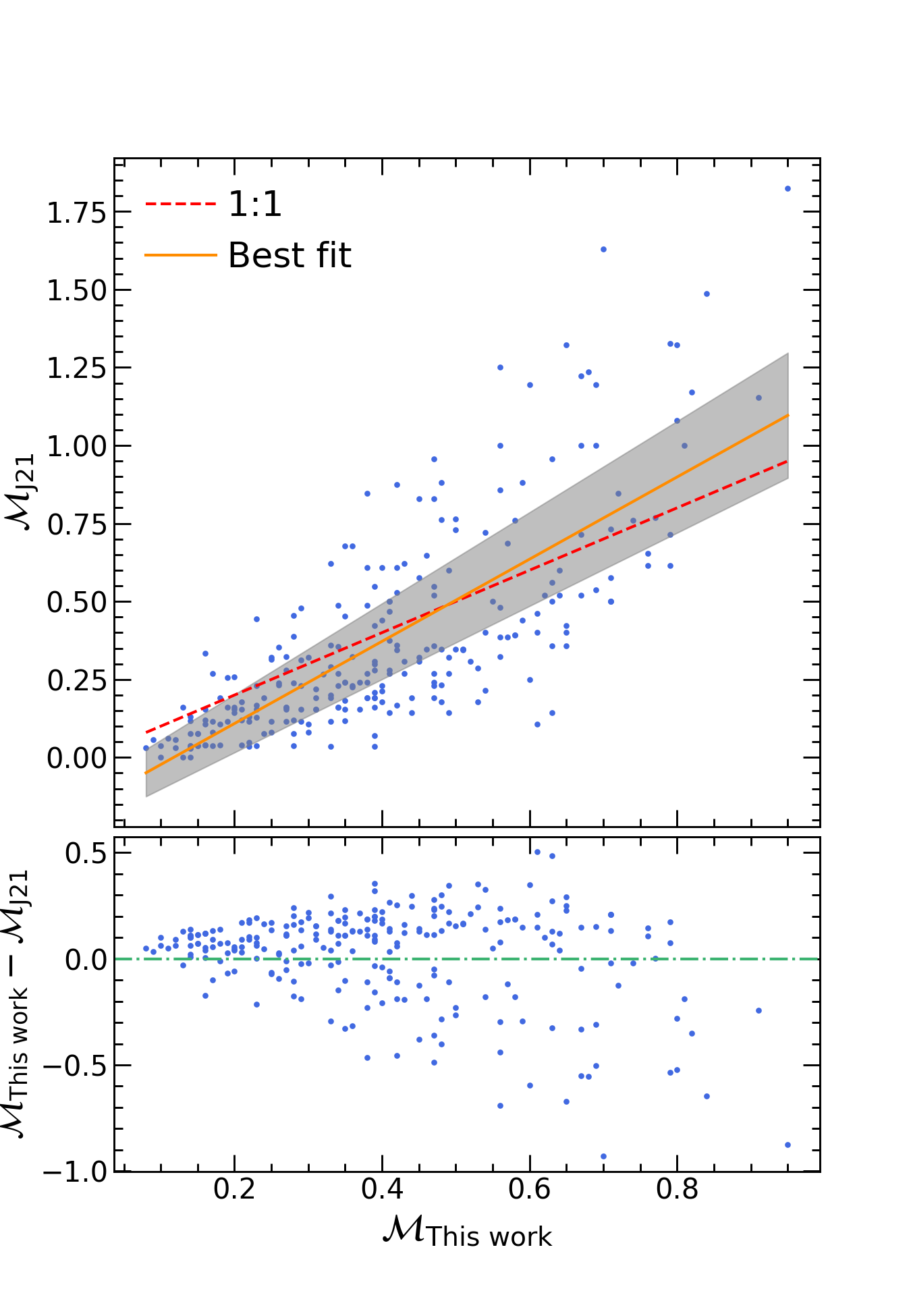}
    \caption{Comparison for $\mathcal{M}_{\rm e}$ obtained values between \citetalias{jadhav_2021} and this work. \textit{Top:} $\mathcal{M}_{\rm e}$ values from \citetalias{jadhav_2021} and our work. 1--to--1 line is shown as a dashed red line and best fit is shown as a solid yellow line. Shaded are shows the $95\%$ confidence interval for the fit. \textit{Bottom:} Difference between \citetalias{jadhav_2021} $\mathcal{M}_{\rm e}$ obtained values vs. $\mathcal{M}_{\rm e}$ found in our work. Green dot-dashed horizontal line denotes $\mathcal{M}_{\rm e, \ This \ work} = \mathcal{M}_{\rm e, \ J21}$.}
    \label{fig:jadhav_comparison}
\end{figure}

In \autoref{fig:jadhav_comparison}, we compare our measured $\mathcal{M}_{\rm e}$ values against those from \citetalias{jadhav_2021}. In the upper panel, the dashed red-line shows the 1-to-1 relation, whereas the yellow solid line marks the best linear fit, with the shaded area representing the $95\%$ confidence interval. Overall, we point out that our values are slightly larger compared with those provided by \citetalias{jadhav_2021}, as can be appreciated in the inferior panel of Figure~\ref{fig:jadhav_comparison}. An explanation is our Gaia DR3 selection criteria, due to the $ 10.5 \ {\rm mag} \leq {\rm G} \leq 19.5 \ {\rm mag}$ magnitude selection. We skipped too bright/saturated stars to avoid systematics, which may have been classified as more massive (very bright) BSSs by our models, especially in OCs. In addition, the systematics slope and scatter in Fig.~\ref{fig:jadhav_comparison} are in part due to \citetalias{jadhav_2021} measuring BSS masses using a different isochrone approximation from ours, i.e.~for every BSS they estimate its magnitude position on the ZAMS and compute its mass based on this model, not on the best-fit isochrone like in our approach.

We recall that \citetalias{jadhav_2021} classified BSSs into three classes, i.e.~low--$\mathcal{M}_{\rm e}$ ($\mathcal{M}_{\rm e} < 0.5$), high--$\mathcal{M}_{\rm e}$ ($0.5 < \mathcal{M}_{\rm e} < 1.0$) and extreme--$\mathcal{M}_{\rm e}$ ($\mathcal{M}_{\rm e} > 1.0$) objects, which are speculated to predominantly form via MT, stellar collisions, and multiple-mergers/MT, respectively. In addition to our direct comparison of OCs and GCs, the difference between our results and those provided by \citetalias{jadhav_2021} is that we were unable to find any stars with $\mathcal{M}_{\rm e} > 1.0$ in the extreme--$\mathcal{M}_{\rm e}$ regime, most likely due to our Gaia photometry cuts. It is important to mention, and as \citetalias{jadhav_2021} have warned, that the ``fractional mass excess'' $\mathcal{M}_{\rm e}$ does not strictly indicate how BSSs are formed, but it rather \emph{suggests} how BSSs may likely be formed. For a deeper analysis, which escapes the Gaia limitations, we do need FUV observations to be able to distinguish the presence or absence of hot companions \citep[e.g.][]{subramaniam_2016, sindhu_2019}.

\begin{figure}
    \includegraphics[width=\columnwidth]{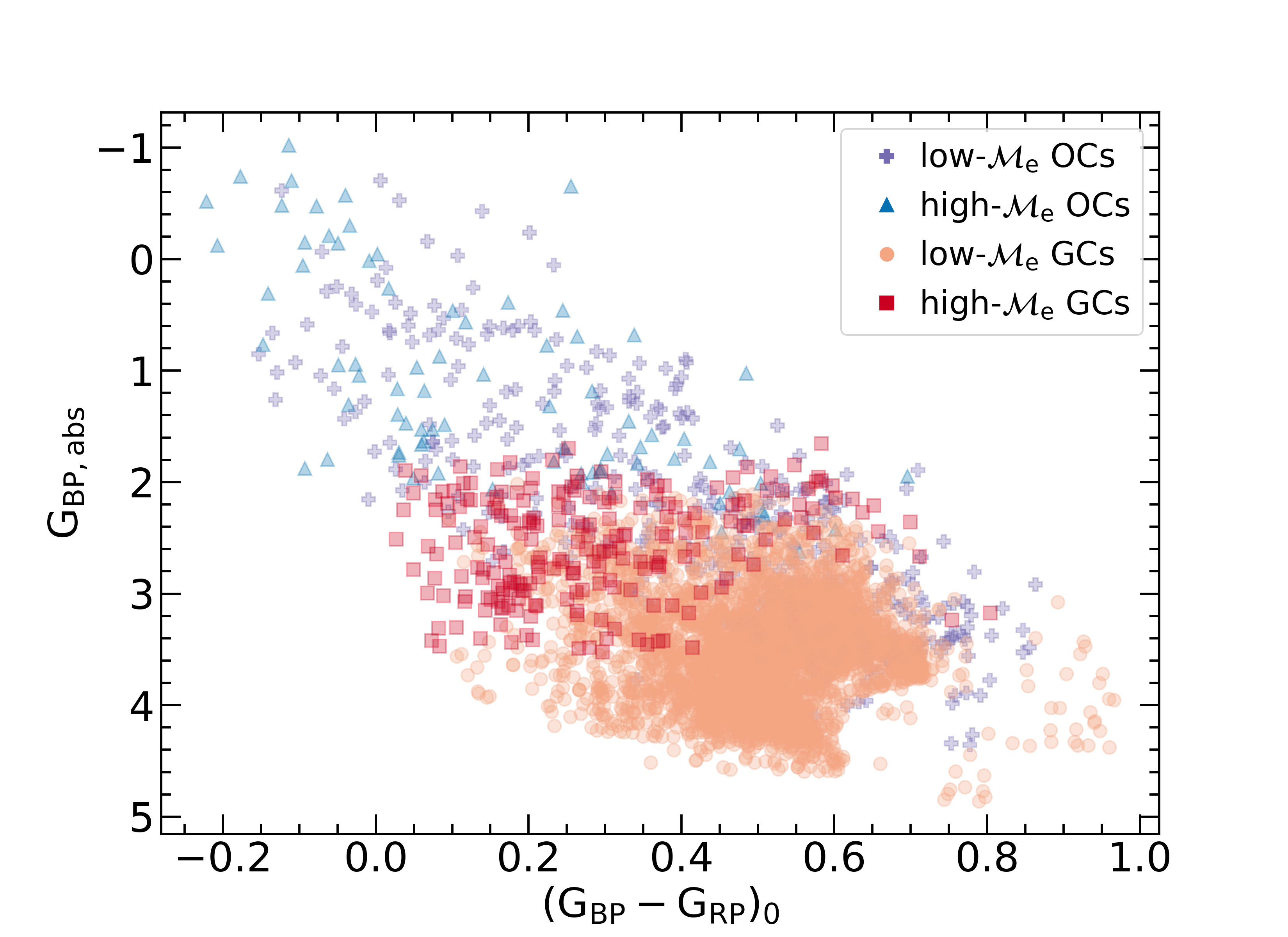}
    \caption{Gaia photometry color-magnitude diagram for BSSs classified by their $\mathcal{M}_{\rm e}$ based on the \citetalias{jadhav_2021} classification. Note that axes show the absolute magnitude for ${\rm G}_{\rm BP}$ and the de-reddened color $({\rm G}_{\rm BP} - {\rm G}_{\rm RP})_0$. Low--$\mathcal{M}_{\rm e}$ BSSs for OCs and GCs are represented by purple plus (\texttt{+}) symbol and orange dots, respectively. High--$\mathcal{M}_{\rm e}$ for OCs and GCs are represented by blue triangles and red squares, respectively.}
    \label{fig:CMD_BSS_classified_by_Me}
\end{figure}

We plot the distribution of low- and high--$\mathcal{M}_{\rm e}$ BSSs in OCs and GCs in the Gaia CMD shown in Figure~\ref{fig:CMD_BSS_classified_by_Me}, similar to \citetalias{jadhav_2021} (see their Fig.~6). We find that BSSs in OCs and GCs classified as high--$\mathcal{M}_{\rm e}$ show a modest difference in color, being bluer than low--$\mathcal{M}_{\rm e}$ BSSs by $\sim\!0.2\,{\rm mag}$. The more significant difference is in the absolute magnitude where high--$\mathcal{M}_{\rm e}$ BSSs are, on average, approximately 10 times brighter than low--$\mathcal{M}_{\rm e}$ BSSs. This difference is appreciable for both types of clusters, GCs and OCs. Although this empirical classification resembles the instantaneous mass excess with respect to the MSTO, it does not resemble the classic evolutionary pathways of stars, which leads them from the MS to the colder subgiant branch and then up the RGB.

\subsection{Toward evolutionary pathways of BSSs} 
\begin{figure}
    \includegraphics[width=\columnwidth]{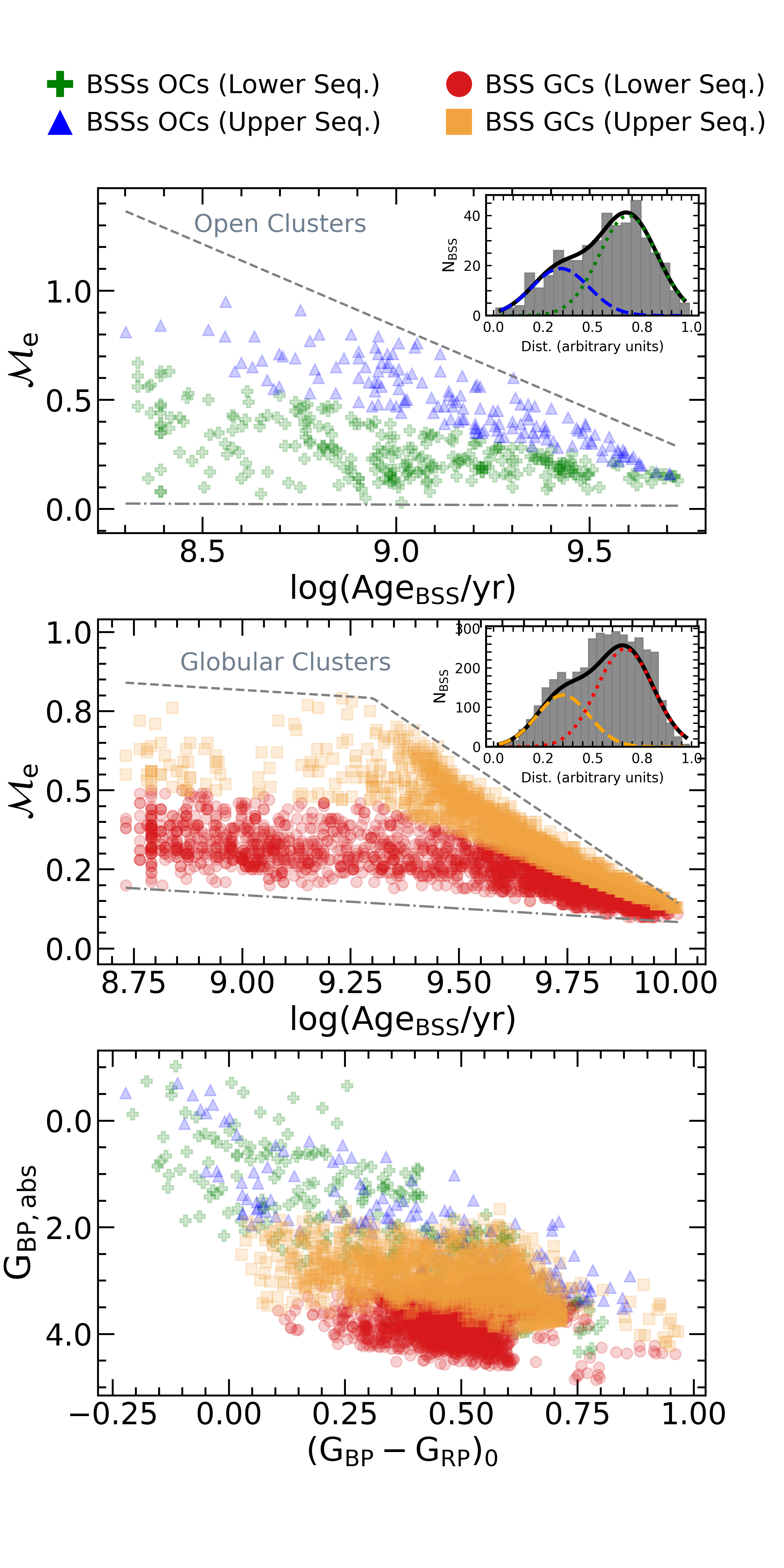}
    \caption{\textit{Top:} $\mathcal{M}_{\rm e}$ vs. BSS age (found with isochrone models) for OCs. \textit{Middle:} $\mathcal{M}_{\rm e}$ vs. BSS age (found with isochrone models) for GCs. In both Top and Middle panel, the dashed gray lines used to separate the BSSs classified as ``upper'' or ``lower'' sequences. OC BSSs and GC BSS that are tagged as ``upper'' sequence BSSs are symbolized by blue triangles and yellow squares, respectively.  ``lower'' sequence for OCs and GCs are represented by green plus (\texttt{+}) symbols and red dots, respectively. \textit{Bottom:} CMD (using absolute magnitudes and de-reddened colors) based on BSSs sequence tag. Symbols are the same as the top figure.}
    \label{fig:upper_lower_Me_classification}
\end{figure}

With the goal of advancing our understanding of the BSS formation channels and capturing the predominance of the BSS formation channels, we combine the fractional mass excess parameter (${M}_{\rm e}$) with our earlier derived equivalent BSS ages. We focus our attention on Figure~\ref{fig:upper_lower_Me_classification}, the upper panel of which reproduces the bottom right panel of Figure \ref{fig:bss_vs_mass}, but this time, we separate the two sequences\footnote{Do not confuse this with the BSS double sequences in a CMD discussed earlier, as the upper panel Figure~\ref{fig:upper_lower_Me_classification} is not a CMD.} into a high-mass excess sequence and low-mass excess through the next procedure:
\begin{enumerate}
    \item For the identified BSSs, we separate them using as criteria if they belong to an OC or a GC and make a plot $\mathcal{M}_{\rm e}$ versus $\log({\rm Age}_{\rm BSS}/{\rm yr})$ (similar as the bottom-right panel of Figure \ref{fig:bss_vs_mass}).
    \item We then set two edges by eye that enclose all the points in this plane. We define an inferior and superior edge, which is different for OCs and GCs. For OCs we define the lower edge defined as:
    \begin{equation}\label{eq:inferior_edge_OCs}
        \mathcal{M}_{\rm e, inf, OC}(x) = -0.007 x + 0.084 
    \end{equation}
    where $x = \log({\rm Age}_{\rm BSS}/{\rm yr})$ is the BSS age found with isochrone fittings. We then define the superior edge for OCs as
    \begin{equation}\label{eq:superior_edge_OCs}
        \mathcal{M}_{\rm e, sup, OC}(x) = -0.756 x + 7.64.
    \end{equation}
    In a similar way, we computed these edges for GCs as
    \begin{equation}\label{eq:inferior_edge_GCs}
        \mathcal{M}_{\rm e, inf, GC} (x) = -0.085 x + 0.934 
    \end{equation}
    and
    \begin{equation}
\mathcal{M}_{\rm e, sup, GC} = 
\begin{cases}
-0.085x + 1.582 & \text{if } x < 9.3 \\
-0.92x + 9.348 & \text{if } x \geq 9.3.
\end{cases}
\end{equation}
   
\item To check the distribution along the horizontal axis, for every point $j$ from OCs in this plane with coordinates $(x_j, \mathcal{M}_{\rm e,j})$, we compute the distance between the point and its superior edge along the vertical axis and normalize this value by the distance between the superior and inferior edges to ``flatten'' it. We can reach this using the expression
\begin{equation}\label{eq:total_distance}
    d(x_j) = \frac{|\mathcal{M}_{\rm e, sup}(x_j) - \mathcal{M}_{\rm e,j}|}{|\mathcal{M}_{\rm e, sup}(x_j) - \mathcal{M}_{\rm e, inf}(x_j)|}.
\end{equation}

\item We repeated the last step, but this time for BSSs from GCs using the respective edges values from step 2.

\item We fit a bimodal distribution using \texttt{SciPy} \citep{scipy_2020} to the distance distribution obtained with equation \eqref{eq:total_distance} for OCs and GCs, separately.

\item Based on the values that optimize the bimodal distribution for the discovered distances, we compute the probability density function (PDF) of each Gaussian and estimate the membership probability of each point to be a member of each Gaussian.

\item If the point has a higher probability to belong to the Gaussian with lower distances, this means that the point belongs to the distribution that is closer to the superior edge, and it is therefore classified as a member of the ``upper'' $\mathcal{M}_{\rm e}$ sequence. Else, they are classified as a member of the ``lower'' $\mathcal{M}_{\rm e}$ sequence.
\end{enumerate}

To properly compare BSSs with this new classification scheme, we plot them in a Gaia photometry CMD in the lower panel of Figure~\ref{fig:upper_lower_Me_classification} (bottom panel), which can be directly compared to Figure~\ref{fig:CMD_BSS_classified_by_Me} which encodes the $\mathcal{M}_{\rm e}$ classification from  \citetalias{jadhav_2021}. Our new classification now resembles more the expected stellar evolutionary pathways of BSSs, clearly separating the BSS population in GCs into two dominant sequences. We also observe that the upper-$\mathcal{M}_{\rm e}$ sequence BSSs in OCs predominantly populate the brighter sequence of OC BSSs in the CMD. In general, we find that the color range covered by the two $\mathcal{M}_{\rm e}$ sequences is similar for GCs and OCs, but there is a systematic offset in luminosity with the upper-$\mathcal{M}_{\rm e}$ sequence BSSs appearing about $1~{\rm mag}$ brighter than their lower-$\mathcal{M}_{\rm e}$ sequence counterparts.

In the following, we briefly explore possible explanations and considerations when interpreting these clearly distinct BSS sequences. 

\subsubsection{Mass transfer evolution}
From a binary evolution perspective, the upper-$\mathcal{M}_{\rm e}$ sequence may represent a BSS population that was formed through more efficient MT processes. This would imply more accreted mass and, hence, more fuel on the receiving star, accelerating its evolution \citep{mccrea_1964}. Assuming this scenario and a subsequent common-envelope evolution, the separation between the stars of such a binary system will eventually decrease until the final stages which could play out in two scenarios: i) the cores could merge or ii) one of the cores could be ejected from the system through interactions with envelope and/or other stars. The former scenario is more likely \citep{hurley_2002}. The result of a stellar merger is a massive BSS, which would experience fast subsequent stellar evolution ($\tau \propto M^{-3}$) and be more likely encountered at older BSS ages. We therefore interpret the two sequences in the $\mathcal{M}_{\rm e}$ versus~$\log({\rm Age_{\rm BSS}}/{\rm yr})$ plane as post-merger/close-binary interaction sequence (``upper'' sequence) and pre-merger/close-binary interaction sequence (``lower'' sequence, see Fig.~\ref{fig:upper_lower_Me_classification}).

\begin{figure}
	\includegraphics[width=\columnwidth]{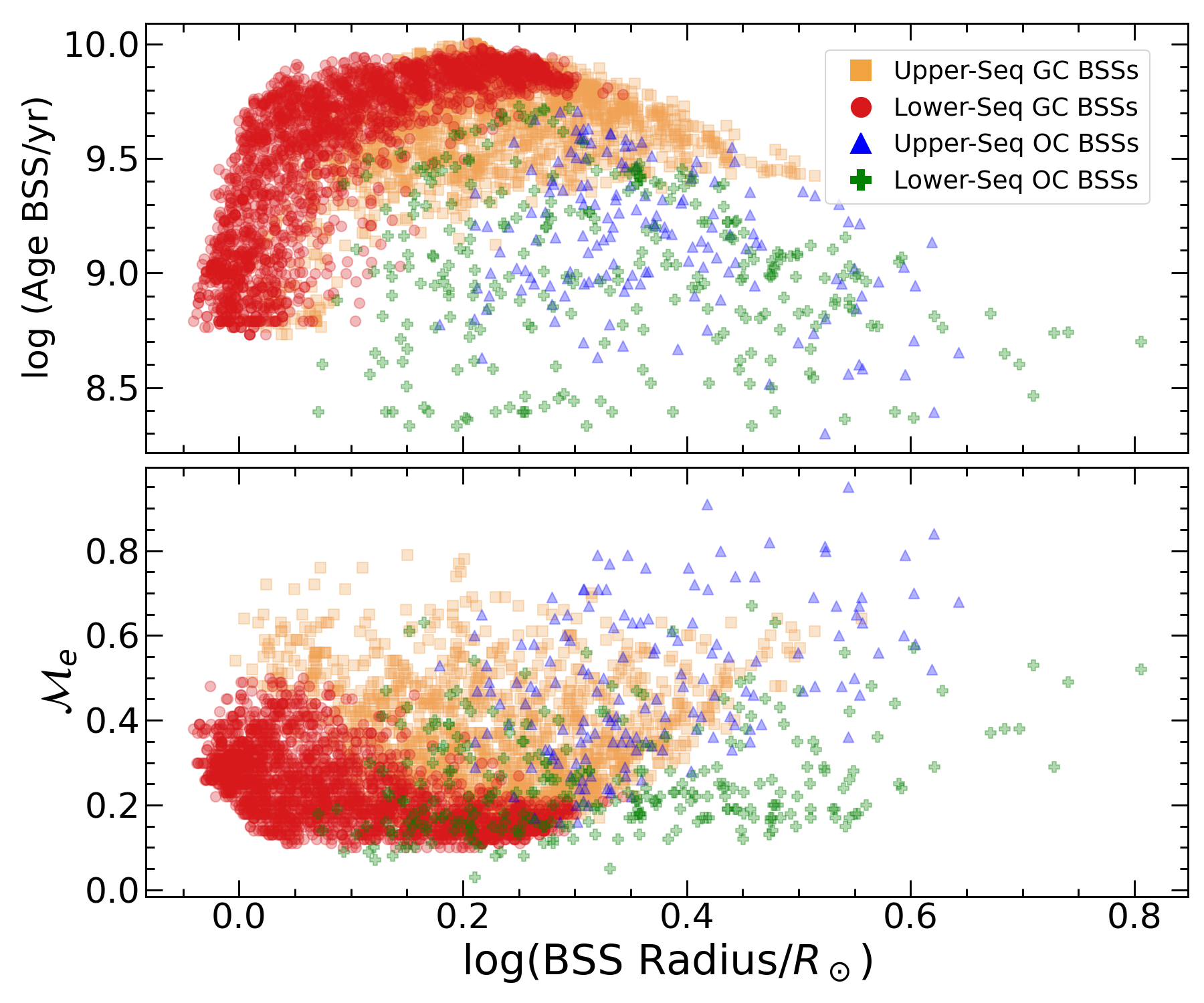}
    \caption{ $\log ({\rm Age}_{\rm BSS})$ and $\mathcal{M}_{\rm e}$ vs. stellar radius. GC upper-sequence, GC lower-sequence, OC upper-sequence and OC lower-sequence BSSs are displayed as orange squares, red dots, blue triangles, and green pluses, respectively.}
    \label{fig:3D_plots}
\end{figure}

\subsubsection{Unresolved binary stars}    
It is trivial that some BSSs may be unresolved binary systems \citep{gosnell_2015, subramaniam_2016, sindhu_2019}. When we fit isochrone models to our BSS sample, however, we implicitly assumed a single-star evolution. \citet{jiang_2022} has shown through multiple simulations that the total mass of the system is higher in unresolved MT binaries compared to post-MT BSSs by about $0.2$--$0.6~M_\odot$. This means that the isochrone fitting method could be overestimating ages and underestimating the masses for unresolved MT BSSs, located preferentially in CMD regions with similar colors as the MSTO and the subgiant branch. Therefore, the upper-$\mathcal{M}_{\rm e}$ sequence could be stars with an overestimated age --showing higher $\mathcal{M}_{\rm e}$ values-- since they are unresolved MT binaries, while the lower-$\mathcal{M}_{\rm e}$ BSSs could be post-MT binaries whose companions either have negligible mass or which are in the post-MT phase.

\subsubsection{Effect of variations in the stellar radius}
As rejuvenated stars evolve of the ZAMS towards the subgiant branch and are therefore expected to grow in size, we compute the radii of our sample BSSs, $R$, using the relation $R\!=\!\sqrt{G M/g}$, where $G$ is the gravitational constant, $M$ is the stellar mass and $g$ is the surface gravity. We assume $g$ in cgs units (from isochrone models) and stellar mass in $M_\odot$ units, computing the value of $R$ in solar radii using \texttt{Astropy} routines \citep{astropy_2013}. We show the results in Figure~\ref{fig:3D_plots}:  $\mathcal{M}_{\rm e}$ and $\log({\rm Age}_{\rm BSS})$ as a function of stellar radius ($R$). We note that after adding stellar radius, the separation between upper- and lower-$\mathcal{M}_{\rm e}$ sequences is even more pronounced for GCs. 

This is consistent with the expectation of an evolved binary system for BSSs in the upper-$\mathcal{M}_{\rm e}$ sequence (orange squares in Fig.~\ref{fig:3D_plots}), where the system might be in the common envelope phase. Such systems evolve rapidly and reach subsequent stages with similar characteristics of an SGB or RGB star, where they begin the radial expansion, which qualitatively agrees with our observations. We were unable to identify any clear difference between the upper- and lower-$\mathcal{M}_{\rm e}$ sequences for BSSs in OCs, however, which may be an indication of a different mix of BSS formation mechanisms acting in lower mass, lower density star clusters. We defer the detailed modeling of these systematic differences between OCs and GCs to future work.

%________________________________________________________________

\section{Conclusions}\label{sec:conclusion}
 
\begin{figure*}
	\includegraphics[width=\columnwidth]{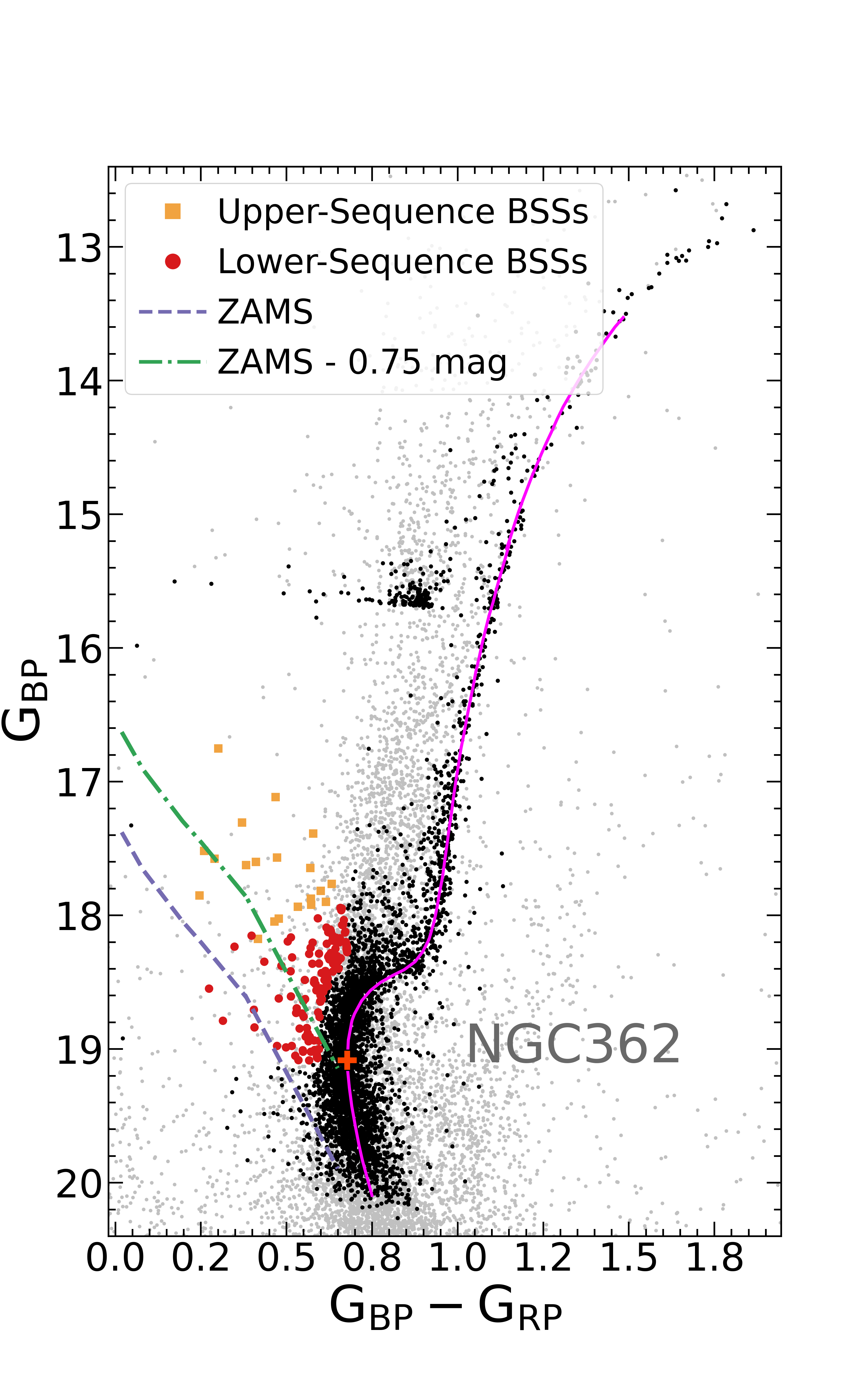}
    \includegraphics[width=\columnwidth]{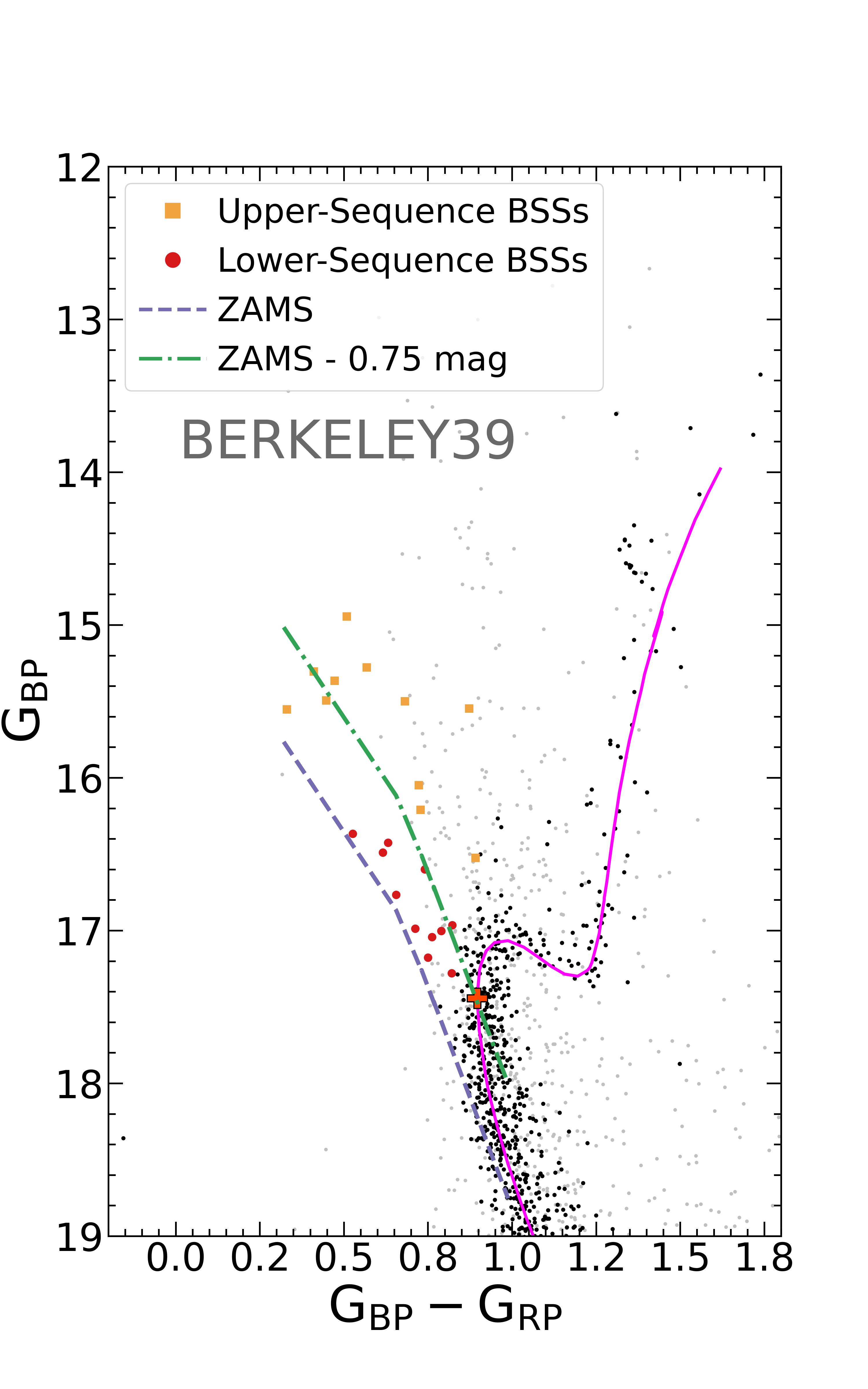}
    \caption{CMD for BSSs classified as ``upper'' and ``lower'' sequences, based on sequences shown Figure \ref{fig:upper_lower_Me_classification}, for GC NGC362 and OC Berkeley 39. ZAMS and ${\rm ZAMS} - 0.75 \ {\rm mags}$ are displayed as a purple dashed line and as a green dash-dotted line, respectively. Orange plus (\texttt{+} symbol) and magenta line represent the MSTO and the best-fitting PARSEC isochrone, respectively (similar to Figure \ref{fig:bss_selection_example}). Field/non-member stars are displayed as gray points, whereas confirmed members are displayed as black points. BSSs that lie in the upper-sequence and lower--sequence are symbolized as orange squares and red dots, respectively.}
    \label{fig:upper_lower_location}
\end{figure*}

With the new $\mathcal{M}_{\rm e}$-$\log({\rm Age_{\rm BSS}}/{\rm yr})$ classification scheme of BSSs, we inspected the CMDs of all our sample GCs and OCs individually to determine whether the BSS subpopulations are differentiated by our classification in the color-magnitude space. We observed that  BSSs classified as upper-$\mathcal{M}_{\rm e}$ members are brighter and that lower-$\mathcal{M}_{\rm e}$ sequence member stars appear fainter. 
In Figure~\ref{fig:upper_lower_location} we presented the distribution of upper- and lower-$\mathcal{M}_{\rm e}$ sequence BSSs in the representative GC NGC\,362 and OC Berkeley\,39. We also overlaid isochrones corresponding to the ZAMS and binary ZAMS. The figure shows that binary ZAMS passes through the majority of the BSSs that intersect the upper- and lower-$\mathcal{M}_{\rm e}$ sequence, while the ZAMS passes through a smaller fraction of the lower-$\mathcal{M}_{\rm e}$ sequence (red points). The lack of BSSs on the ZAMS also suggests the lack of collisional BSSs, which is also expected because NGC\,362 is a core-collapse GC and the Gaia data are unable to resolve the core of the GCs. We do not expect to observe collisional BSSs in OCs either because the environment is not dense enough to favor collision. 
Our classification is not similar to the literature classification of blue and red BSSs sequences observed in core-collapse GCs, where blue BSSs are thought to form through direct collisions and the BSSs in the red sequence are thought to form via MT \citep{ferraro_2009, dalessandro_2013}. We note that blue (or collisional) BSSs are dominated by our lower sequence, whereas the red sequence is a mix of the upper and lower sequence. Our new BSSs classification based on $\mathcal{M}_{\rm e}$-$\log({\rm Age_{\rm BSS}}/{\rm yr})$ mainly highlights the pre- and post- merger/MT scenarios of the BSSs. Our study suggests that companions in our $\mathcal{M}_{\rm e}$-$\log({\rm Age_{\rm BSS}}/{\rm yr})$ classification may imply that most of our stars that are classified as upper-$\mathcal{M}_{\rm e}$ sequence BSSs might be consistent with the presence of a companion, while lower-$\mathcal{M}_{\rm e}$ BSSs may preferentially indicate post-MT/merger-product BSSs. This may support our hypothesis, which assumes that we might be overestimating the ages of BSSs because we cannot resolve the binary system, and therefore, the BSSs were classified as upper-$\mathcal{M}_{\rm e}$ sequence members. 

The distribution of upper- and lower-$\mathcal{M}_{\rm e}$ sequence BSSs in the CMDs for GCs and OCs is similar for all clusters. The CMDs of all the clusters highlighted with these two sequences are made available on \href{https://zenodo.org/records/15202637}{Zenodo} for reference. In some of the clusters (NGC 104, NGC288, NGC 2808, NGC 3201, NGC 5139, and NGC 6121) that are not classified as core-collapse GCs, we observed that a significant number of BSSs are located on the ZAMS or are bluer than the ZAMS. These most probably formed through a direct collision or post-MT/merger, as suggested by the literature on core-collapse GCs. The significant number of BSSs outside the core or half-light radius of the clusters makes these GCs interesting candidates and requires a further detailed analysis to interpret the observed nature.

Much more theoretical work and observations are required to elucidate whether and to which extent our new classification scheme of BSSs resembles the formation and evolution mechanisms of these unique stars. We summarize the results and conclusions of our work below.

\begin{enumerate}
    \item We selected OCs from \citet{dias_2021} with more than 350 star members and GCs from \citet{vasiliev_2021} with more than 1000 star members to ensure well-sampled cluster stellar populations, and we used data from Gaia DR3, for which we adopted an algorithm based on prescriptions from \citetalias{cordoni_2018} and \citet{cordoni_2020} to extract cluster members. 
    
    \item Using the confirmed cluster members, we fit PARSEC isochrones \citep{bressan_2012} to obtain parameters such as the age for every cluster in our sample.
    
    \item Using the ridge line of the observed CMD and the best-fit isochrone found in the previous step along with the ZAMS and the dispersion of stars along these sequences, we defined a BSS selection. We were able to find BSSs in all our sample GCs and in 42 out of 129 ($\sim\!33\%$) OCs of our sample. We identified a total of 4399 BSSs, 434 of which ($\sim\!10\%$) are located in OCs and 3965 ($90\%$) are located in GCs.
   
    \item We extracted astrophysical parameters for all identified BSSs using three different methods: 1) color--temperature relations (i.e., IRFM), 2) model isochrone fitting, and 3) parameters from Gaia DR3 spectra from \citet[][called XPP]{zhang_2023}, if available. From the color--temperature relations, we only obtained $T_{\rm eff}$. From the isochrone fitting, we obtained $T_{\rm eff}$, $\log g$, $\log L$, stellar masses, and ages. From \texttt{XPP} we obtained $T_{\rm eff}$ and $\log g$. 

    \item We found no BSSs in clusters younger than $\sim\!500\,{\rm Myr}$.
    
    \item We found effective temperatures for BSSs $\langle T_{\rm eff}\rangle = 6853 \pm 795\,{\rm K}$ from the color--temperature relations and $\langle T_{\rm eff}\rangle = 6900 \pm 687\,{\rm K}$ from the isochrone fitting for our entire BSS sample. The $T_{\rm eff}$ found in OC BSSs are $\langle T_{\rm eff}\rangle = 7598 \pm 1476\,{\rm K}$ and $\langle T_{\rm eff}\rangle = 7549 \pm 1331\,{\rm K}$ from the color--temperature and isochrone--fitting methods, respectively. The BSSs in GCs present slightly lower temperatures, with $\langle T_{\rm eff}\rangle = 6772 \pm 630\,{\rm K}$ derived from color--temperature relations and $\langle T_{\rm eff}\rangle = 6829 \pm 529\,{\rm K}$ from isochrone models. Overall, the difference in $T_{\rm eff}$ found from isochrone fittings is that they are $\sim\!50\,{\rm K}$ hotter than those obtained from color--temperature relations. These values agree with those of previous studies.

    \item Based on the isochrone--fitting method, we determined that the average mass of all our BSSs is approximately $\langle M_{\rm BSS} \rangle = 1.09 \pm 0.28\,M_\odot$. The average mass of BSSs in OCs is at $\langle M_{\rm BSS} \rangle = 1.75 \pm 0.45\,M_\odot$. We obtained an average BSS mass for GCs of $\langle M_{\rm BSS} \rangle = 1.02 \pm 0.1\,M_\odot$. Our average mass for BSSs in OCs is lower than the mass determined by \citetalias{jadhav_2021}, which is likely due to our more conservative selection, but they still agree within their respective error ranges. Our results agree very well with most other studies.

    \item We compared $T_{\rm eff}$ against its counterpart measured at the MSTO of the parent cluster. Regardless of the method we applied, we found for OCs and GCs that the difference between the mean BSS $T_{\rm eff}$ and the MSTO $T_{\rm eff}$ decreases with the cluster age. This is in line with expectations from stellar evolution because massive stars evolve faster than stars at the MSTO. The $T_{\rm eff}$ dispersion in OCs decreases with the cluster age. We did not find this last relation in GCs, which means that the $T_{\rm eff}$ dispersion of BSSs does not depend on the cluster age in GCs.
    
    \item We followed the definition provided by \citetalias{jadhav_2021} for the fractional mass excess, or $\mathcal{M}_{\rm e}$. This parameter quantifies the difference between the BSS mass and the stellar mass at the MSTO (this last based on isochrone fittings), normalized by the MSTO mass. $\mathcal{M}_{\rm e}$ is roughly equal or proportional to the MT efficiency in a binary system. These authors classified $\mathcal{M}_{\rm e}$ into three regimes, each assumed to be dominated by a specific BSS formation pathway: i) low--$\mathcal{M}_{\rm e}$ ($\mathcal{M}_{\rm e} < 0.5$, likely to be formed via MT), high--$\mathcal{M}_{\rm e}$ ($0.5 < \mathcal{M}_{\rm e} < 1.0$, likely to be formed through mergers or collisions), and extreme--$\mathcal{M}_{\rm e}$ ($1.0 < \mathcal{M}_{\rm e} < 1.5$, likely to be multiple MT systems). In our sample, we found no extreme--$\mathcal{M}_{\rm e}$, which is likely to be due to the conservative selection, where we avoided stars brighter than $G_{\rm BP}\lesssim10$ (to avoid Gaia systematic errors), which may contain some more massive BSSs. In our final BSS sample, $7\%$ are classified as high--$\mathcal{M}_{\rm e}$ and $93\%$ as low--$\mathcal{M}_{\rm e}$. This proportion is $\sim\!81.3\%$ and $\sim\!18.7\%$ for low--$\mathcal{M}_{\rm e}$ and high--$\mathcal{M}_{\rm e}$, respectively, for OCs. For BSSs in GCs, we found these percentages to be $\sim\!94.3\%$ for low--$\mathcal{M}_{\rm e}$ and $\sim 5.7\%$ for high--$\mathcal{M}_{\rm e}$. This suggests that mass transfer is the most likely formation channel for BSSs, but OCs show a higher fraction of high--$\mathcal{M}_{\rm e}$ BSSs, some of which may have formed through mergers as well.

    \item Finally, we explored a new BSS classification scheme that uses $\mathcal{M}_{\rm e}$ versus $\log({\rm Age_{\rm BSS}})$. This parameter space shows a clear double sequence for BSS in GCs: One parameter space corresponds to an upper sequence, with higher $\mathcal{M}_{\rm e}$ as a function of the BSS age, and another lower sequence that is linked to lower $\mathcal{M}_{\rm e}$ as a function of the BSS age. We plotted the BSSs labeled according to this classification scheme in the Gaia CMD and found a clear division between these branches, especially for GCs. BSSs on the upper-$\mathcal{M}_{\rm e}$ sequence are brighter than those located on the lower-$\mathcal{M}_{\rm e}$ sequence. We explored two possible explanations: 1) In binary star evolution, mass transfer increases the mass of the receiving star and in the process decreases the distance between the binaries, while the receiving star begins an accelerated evolution and therefore appear as an older star in the isochrone tracks. In this scheme, these upper- and lower-$\mathcal{M}_{\rm e}$ sequences are consistent with post-merger or close-binary interaction and pre-merger binary evolution. 2) We might be overestimating the age and underestimating the mass of the upper-$\mathcal{M}_{\rm e}$ sequence BSSs because we assumed single-star evolution during the isochrone fitting. Therefore, the upper- and lower-$\mathcal{M}_{\rm e}$ sequences may be equivalent to an overestimated and realistic BSSs age, which could be correlated with the presence (or absence) of a BSS companion. Further observations targeting various parameters, such as the stellar rotation speed statistics, chemical compositions, and binarity, are required to shed light on this intriguing new bifurcation in the $\mathcal{M}_{\rm e}$ versus $\log({\rm Age_{\rm BSS}})$ classification.
\end{enumerate}

\section{Data availability}
% Parameters of BSSs, YSSs, RSSs and the star clusters presented in this study are made available at the CDS and on \href{https://zenodo.org/records/15202637}{Zenodo} for reference. 

Parameters of BSSs, YSSs, RSSs and the star clusters presented in this study (full Tables D.1 and D.2) are available at the CDS via \href{https://cdsarc.cds.unistra.fr/viz-bin/cat/J/A+A/699/A142}{https://cdsarc.cds.unistra.fr/viz-bin/cat/J/A+A/699/A142} and on \href{https://zenodo.org/records/15202637}{Zenodo} for reference. The CMDs for all the clusters as shown in \autoref{fig:upper_lower_location} are also made available on \href{https://zenodo.org/records/15202637}{Zenodo}.

\begin{acknowledgements}
      P.K.N. and T.H.P. acknowledge support by the ANID BASAL projects FB210003. T.H.P. gratefully acknowledges support through the FONDECYT Regular Project No.~1201016.
\end{acknowledgements}

   \bibliographystyle{aa} % style aa.bst
   % \bibliography{biblio} % your references Yourfile.bib
   \bibliography{BSS-GaiaDR3-Arxiv-v2} % your references Yourfile.bib
%-------------------------------------------------------------------

\begin{appendix}

\section{Validation of our membership criteria:}
\label{membership_validation}
\begin{figure}
    \includegraphics[width=0.82\columnwidth]{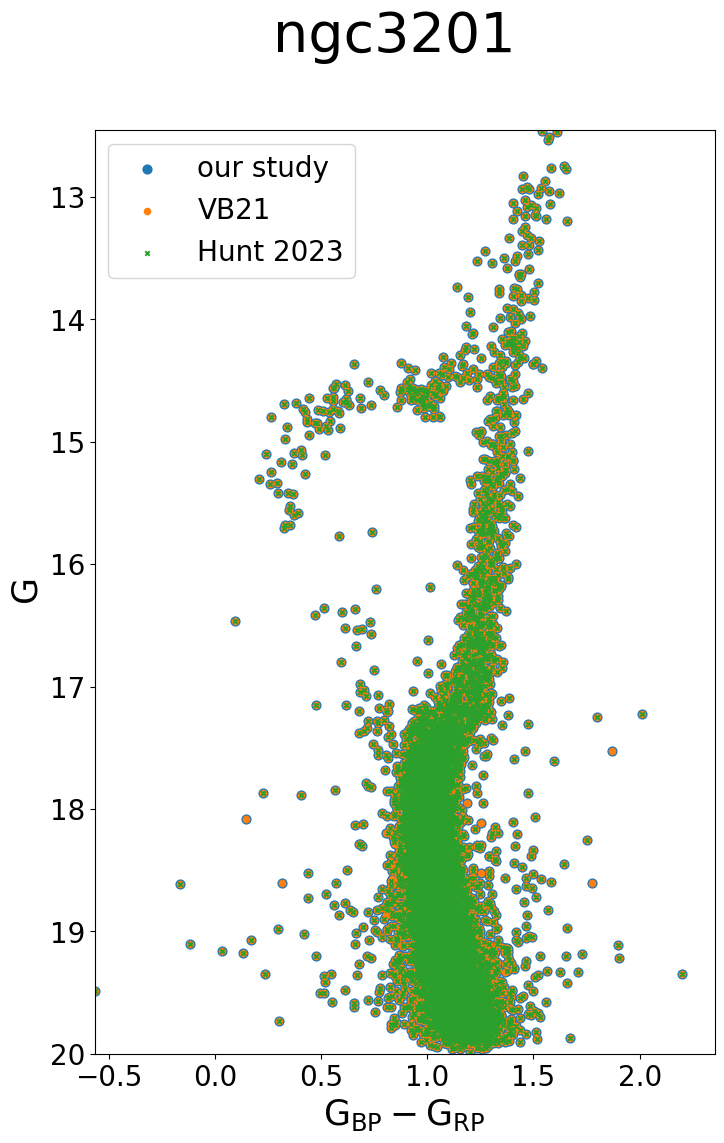}
    \includegraphics[width=0.82\columnwidth]{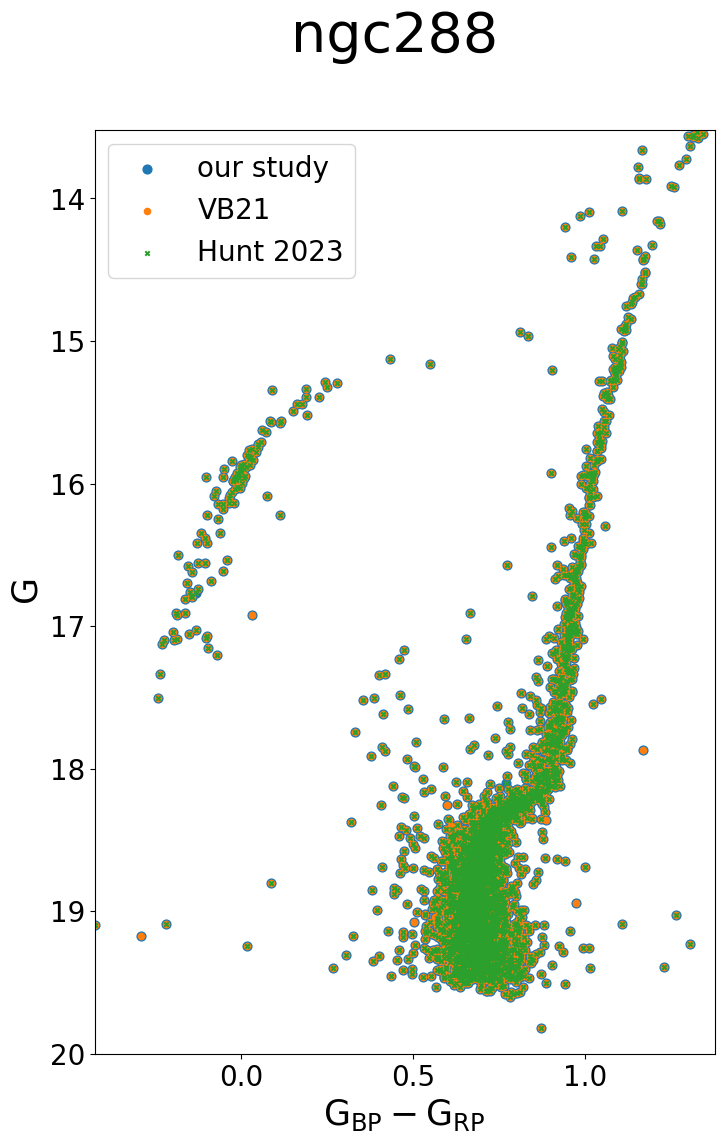} 
    \caption{Comparison of our cluster membership with literature. Two GCs are compared, NGC 3201 and NGC 288. We define \citet{vasiliev_2021} as VB21; \citet{Hunt_2023_OCs} as Hunt 2023. 
}
    \label{fig:compare_membership_method_GC}
\end{figure}

To validate our BSS membership criteria as well as all other cluster members, we compare our cluster members with the literature. We only compare the clusters which show the presence of BSSs as we only focus on these clusters in our further analyses. The most recent and extensive membership catalogs using Gaia data are given by \citet{vasiliev_2021, Hunt_2023_OCs} for GCs and \citet{Cantat-Gaudin_2020_OCs, Hunt_2023_OCs} for OCs. \citet{vasiliev_2021} provided membership probability of all stars in 170 GCs using the Gaussian mixture model on Gaia-EDR3 data, where stars with a membership probability greater than 90\% are considered to be cluster members. We find more than 90\% matches in membership for all 41 GCs, with a median crossmatch accuracy of 99.7\%. All BSSs detected in our GCs sample are also classified as cluster members in \citet{vasiliev_2021}. \citet{Hunt_2023_OCs} used HDBSCAN algorithm on Gaia DR3 data down to magnitude G $\sim$ 20 mags to recover 7167 clusters, including 134 globular clusters. 
For GCs, we find that the crossmatch between our catalog and \citet{Hunt_2023_OCs} varies from 15\% (for NGC 6656; very close to the Galactic plane) to $>$95\% based on the distance of the clusters from the Galactic plane, with a median crossmatch accuracy of 84\%. NGC 104 is not studied by \citet{Hunt_2023_OCs}, hence we are unable to make any crossmatch with this cluster. The membership method of \citet{Hunt_2023_OCs} is found to have a bias on whether the cluster is located in a crowded or sparse field, which is not found in the methods of \citet{vasiliev_2021}. In \autoref{fig:compare_membership_method_GC}, we show CMDs of two GCs (NGC 3201 and NGC 288) to demonstrate the comparison of the cluster membership obtained from our method (blue), \citet{vasiliev_2021} in orange and \citet{Hunt_2023_OCs} in green. The figure shows a good agreement between all three methods.

\begin{figure}
    \includegraphics[width=0.82\columnwidth]{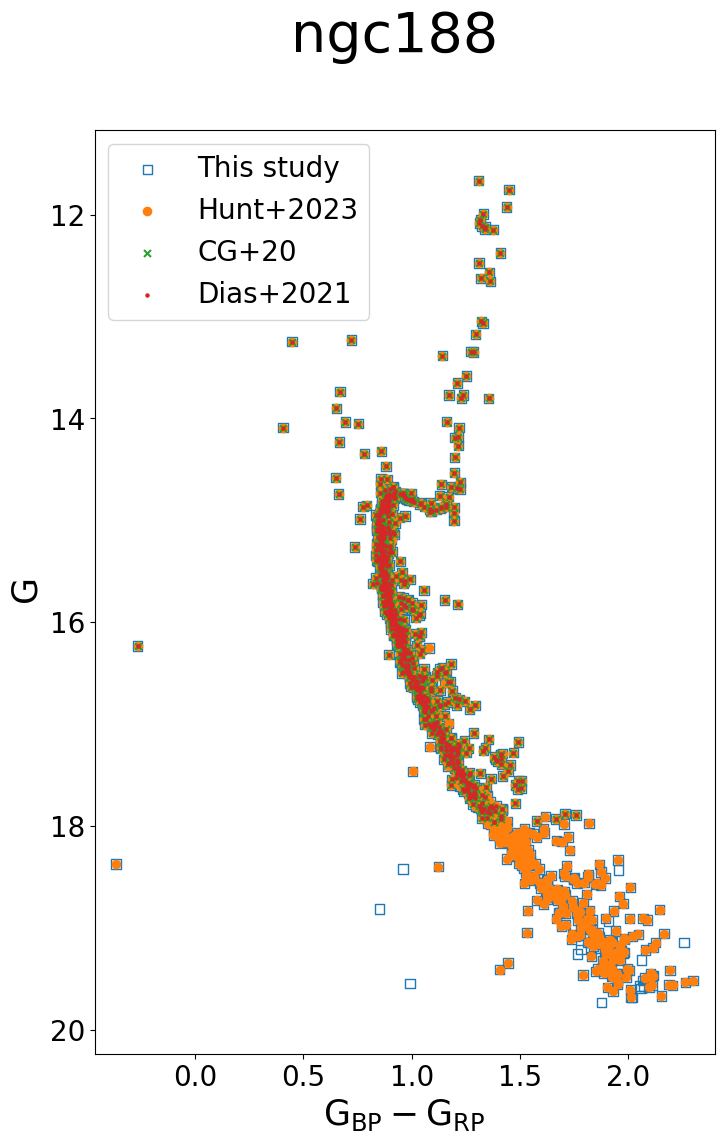}
    \includegraphics[width=0.82\columnwidth]{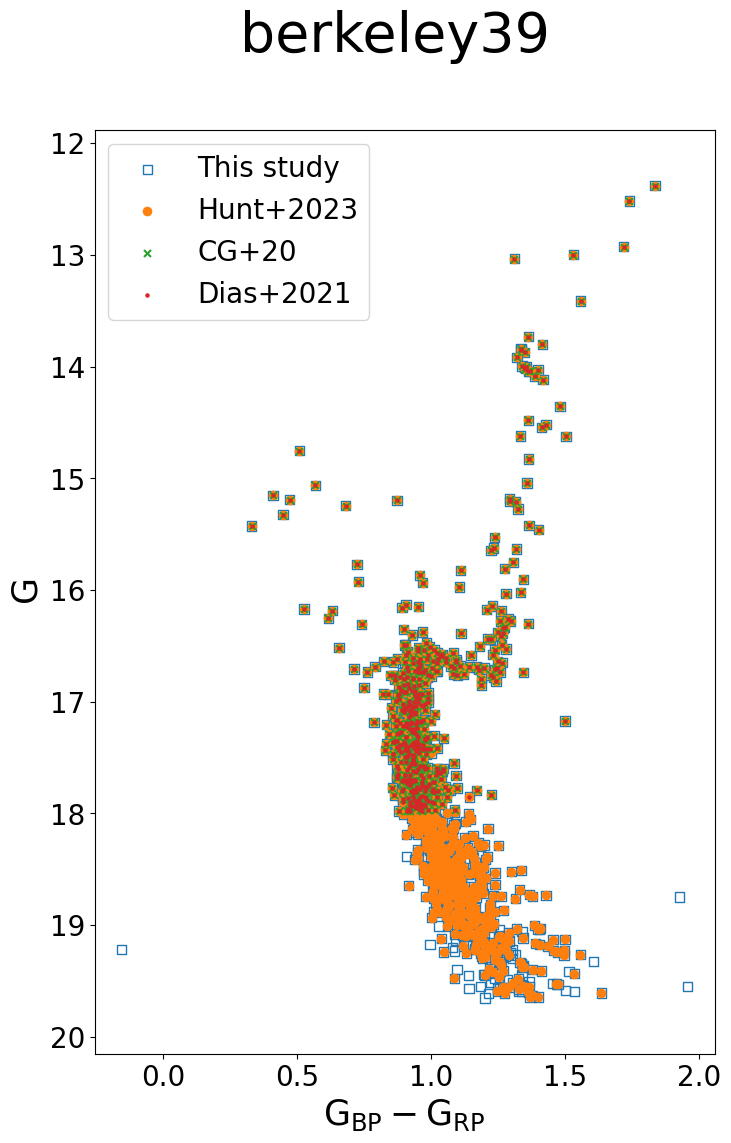}
    \caption{Comparison of our cluster membership with literature. Two OCs are compared, NGC188 and Berkeley 39. We define \citet{Hunt_2023_OCs} as Hunt+2023; \citet{cantat_gaudin_2020} as CG+20 and \citet{dias_2021} as Dias+2021.
}
    \label{fig:compare_membership_method_OC}
\end{figure}

%% ----------------------------------------------

We find a better crossmatch with  \citet{Hunt_2023_OCs} for OCs. We find a crossmatch of more than 50\% for all clusters (except UBC252 with 19\%) and with median crossmatch value of 94\%. The unmatched sources are mainly fainter than $\sim$18 mags in G band and more importantly, all BSSs in our sample are also classified as cluster members in \citet{Hunt_2023_OCs}.  \citet{dias_2021} used the classic maximum likelihood approach \citep{dias_2014, Monteiro_2020} on Gaia-DR2 data to study $\sim$2000 open clusters for stars brighter than 18 mags in the G band. The comparison shows that the crossmatch fraction varies from 60\% to 96\% with a median value of 88\%. The difference in the crossmatch is mainly found towards the fainter G magnitudes. When cross-matched only with our BSS sample, we find more than 97\% are also classified as members in \citet{dias_2021}.   
\citet{Cantat-Gaudin_2020_OCs} applied an artificial neural network in Gaia DR2 photometry down to G = 18 mags to estimate the cluster membership of $\sim$2000 OCs. Comparing our cluster members with stars brighter than 18 mags in G band, we find that crossmatch fraction varies from cluster to cluster with a range from 40\% to 95\%. There are mainly two reasons behind the differences in crossmatch fraction: \citet{Cantat-Gaudin_2020_OCs} used Gaia DR2 for the membership instead of DR3 with better astrometry and was unable to recover the cluster member properly towards the fainter magnitude limit. For example, there are no cluster member present with G$>$17 mags for the cluster UBC 252, which leads to only 41\% of our cluster member. While we consider the match for BSSs, we notice that out of 434 BSSs candidate in our catalog, 409 ($>$94\%) are also part of the cluster member in \citet{Cantat-Gaudin_2020_OCs}. In \autoref{fig:compare_membership_method_OC}, we present CMDs of two OCs (NGC 188 and berkeley 39) as examples to display the comparison between our cluster membership (open blue box) to those obtained by \citet{Hunt_2023_OCs} marked in orange points, \citet{Cantat-Gaudin_2020_OCs} marked in green cross, and \citet{dias_2021} in red points. The cluster memberships are found to be well-matched among the different methods. 
\citet{jadhav_2021} also provides an extensive catalog of BSSs candidates in OCs. They used the primary catalog of OCs by \citet{Cantat-Gaudin_2020_OCs}, however. We therefore did not perform a detailed cross-match with \citet{jadhav_2021}. Therefore, we find that our cluster membership matches well with the literature and almost all BSSs are also classified as members in the literature, which also validates the membership determination algorithm used in our study.

\section{Analysis} \label{sec:analysis}

To obtain astrophysical parameters for BSSs we use three different approaches: i) Color--temperature relations; ii) Model isochrone fitting; iii) Spectroscopic data from Gaia DR3 (if available).

%________________________________________________________________
\begin{figure*}[!ht]
    \includegraphics[width=\textwidth]{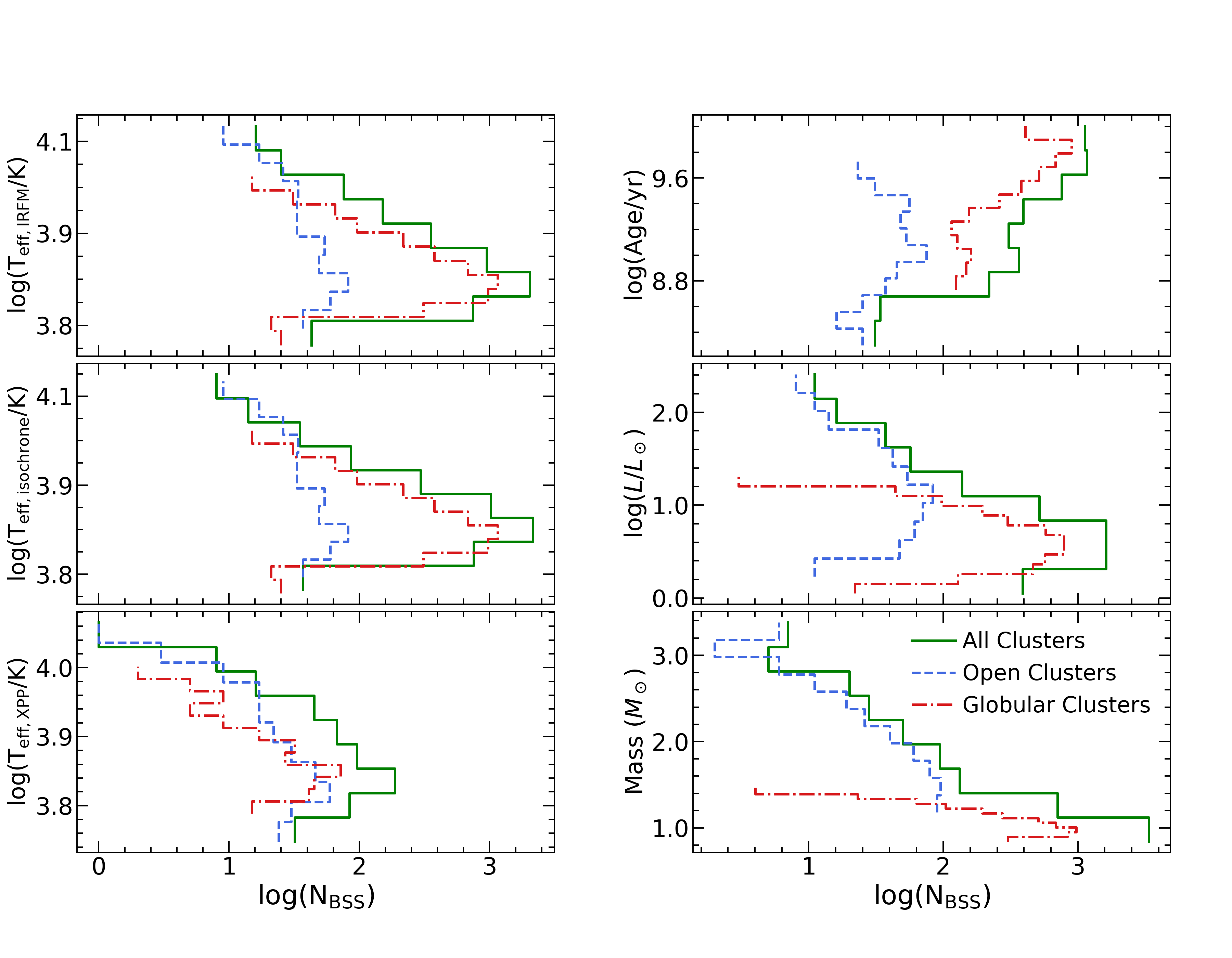}
    \caption{Histogram of derived astrophysical parameters for our sample BSSs. Top, middle and bottom left panels are the distributions for effective temperatures derived from color--temperature relations, isochrone fittings, and the XPP method (from Gaia DR3 spectroscopy, if available), respectively. Top, middle and bottom right panels are the distributions obtained from isochrone fitting, for BSS ages, luminosities and masses, respectively. GCs are denoted as a red dotted--dashed line, OCs as a dashed blue line and the distribution for all clusters (GCs + OCs) is depicted as a solid green line.}
    \label{fig:histogram_teff_and_mass}
\end{figure*}

\begin{figure}
	\includegraphics[width=\columnwidth]{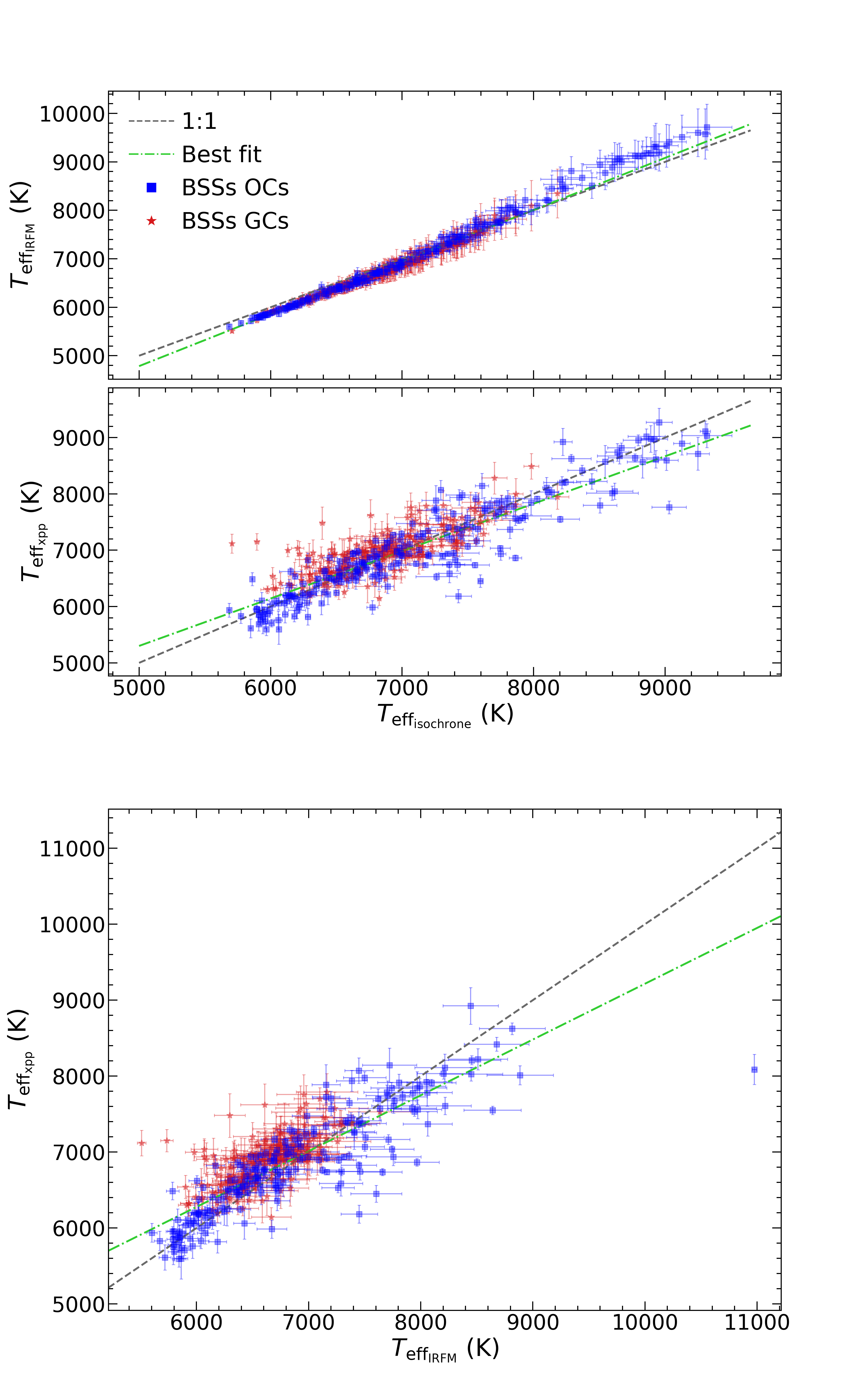}
    \caption{\textit{Top:} $T_{\rm eff}$ obtained with isochrone best-fitting method \citep[from PARSEC isochrones,][]{bressan_2012} compared against $T_{\rm eff}$ obtained via IRFM \citep[color--temperature relations,][upper panel]{gonzalez_2009} and XPP \citep[][lower panel]{zhang_2023}. \textit{Bottom:} $T_{\rm eff}$ obtained with XPP vs. $T_{\rm eff}$ obtained with IRFM. The 1--to--1 relation is shown as a gray dashed line, whereas the best fit (considering GC and OC stars) is shown as a dot-dashed green line. BSSs from GCs are displayed as red stars and BSSs from OCs as blue squares. The figure only displays BSSs whose error in both components is less than $300~{\rm K}$.}
    \label{fig:effective_temperature_relations_methods}
\end{figure}

%--------------------------------------------------

\subsection{Color--temperature relations}
\label{sec:color_temp}

Based on the infrared flux method (IRFM) from \citet{gonzalez_2009}, the effective temperature of a star can be estimated using the following relation.

\begin{equation}\label{eq:color_temperature_mucciarelli}
    \theta = b_0 + b_1 C + b_2 C^2 + b_3 [{\rm Fe}/{\rm H}] + b_4 [{\rm Fe}/{\rm H}]^2 + b_5[{\rm Fe}/{\rm H}]C
\end{equation}
where $\theta = 5040 \ {\rm K}/T_{\rm eff}$, $C$ is the dereddened color and $[{\rm Fe}/{\rm H}]$ is the metallicity. We use the coefficients $b_{i=0-5}$ from \citet{mucciarelli_2021}, which are based on a Gaia EDR3 analysis of dwarf and giant stars. $[{\rm Fe}/{\rm H}]$ are taken from \citet{carretta_2009} and \citet{dias_2021} for GCs and OCs, respectively. For Pismis 3 and Trumpler 23, metallicities were taken from \citet{bisht_2022} and \citet{overbeek_2017}, respectively.

\subsection{Isochrone-fitting models}
\label{sec:iso_method}

Previous studies have attempted to extract BSS parameters based on isochrone models such as \citetalias{jadhav_2021} and \citetalias{dattatrey_2023}. In the first study, the authors assumed that the mass of the BSSs to be approximately the mass of a point at the ZAMS with the same magnitude. In the second study, they fit multiple isochrones with different ages and compare BSS positions with respect to the isochrones. We take an approach similar to the second one. For GCs (OCs), we use multiple PARSEC isochrones \citep{bressan_2012} from 200 Myr (100 Myr) in steps of 50 Myr (25 Myr) up to the cluster age found with the best-isochrone fitting, fixing the metallicity from the literature for each cluster as described earlier. We select the three closest points on the best-fit isochrone for each cluster and compute the average and rms for mass, luminosity, age, surface gravity and effective temperature for each BSS.

\subsection{Spectroscopy from Gaia DR3}
\label{sec:xpp}
Gaia DR3 includes optical low--resolution spectra for over 220 million stars. As a brief summary, these spectra are measured with two instruments. One covers the blue--optical range (380--680 nm, measured with the ``Blue Photometer'' instrument) called \texttt{BP} and another covers the infrared range (640--1050 nm, measured with the ``Red Photometer'') called \texttt{RP} \citep{de_angeli_2023, montegriffo_2023}. These \texttt{BP}/\texttt{RP} spectra, hereafter referred to as XP spectra, have a spectral resolution $R\!\simeq\!50\!-\!160$ with parameters inferred from ``classical'' models \citep[][and references therein]{andrae_2023}, which are highly sensitive to errors in low-resolution spectra, which is the case for Gaia DR3 XP spectra. In another approach, \citet{zhang_2023}, based on parameters obtained from higher-resolution spectra ($R\!\simeq\!1800$) from other surveys, cross-matched them with Gaia DR3 spectra and created a machine-learning model that has the advantage of: i) obtaining astrophysical parameters from higher-resolution spectra, comparing them with the XP spectra and adjust those parameters at lower spectral resolution; ii) being able to find various systematic errors that could not easily be noticed in the XP spectra alone. For this reason, we extract XP parameters (hereafter called XPP) from \citet{zhang_2023} for all our sample BSSs, if available, such as effective temperature and surface gravity. We cross-match the provided parameters with our selected BSSs using \texttt{Gaia Source ID} unique identifier. Finally, \citet{zhang_2023} provides a \texttt{quality\_flag} parameter which, in short, is a number to quantify the reliability of the obtained parameters from spectra. The authors suggest a ``basic reliability cut'' to be at least $\texttt{quality\_flag} < 8$ (note, a lower number indicates more reliable parameters). We decided to only use stars with $\texttt{quality\_flag} \leq 4$ to ensure we extract only reliable results.~More details for these parameters are provided in Section 4 of \citet{zhang_2023}.

From the IRFM we find $\langle {\rm T}_{\rm eff, IRFM}\rangle = (6772 \pm 630)~{\rm K}$ for GCs and $\langle {\rm T}_{\rm eff, IRFM}\rangle = (7598 \pm 1476)~{\rm K}$ for OCs. From isochrone fitting we find $\langle {\rm T}_{\rm eff, isochrone}\rangle = (6829 \pm 529)~{\rm K}$ for GCs and $\langle {\rm T}_{\rm eff, isochrone} \rangle = (7549 \pm 1331)~{\rm K}$ for OCs. Only 537 of 4399 BSSs ($\sim\!12\%$ of our sample) had XPP parameters measured for 279 stars in GCs and 258 stars in OCs. With this method, we find the mean BSS effective temperature of $\langle {\rm T}_{\rm eff, XPP}\rangle = (7271 \pm 732)~{\rm K}$ for GCs and $\langle {\rm T}_{\rm eff, XPP} \rangle = (7269 \pm 1331)~{\rm K}$ for OCs. Figure~\ref{fig:histogram_teff_and_mass} shows the distribution for $T_{\rm eff}$, $L$, and $M$ of all our sample BSSs. 
We also estimated the parameters for YSSs and RSSs, which are available at the CDS.
Their distributions are shown in \autoref{fig:histogram_teff_and_mass_yss} and \autoref{fig:histogram_teff_and_mass_rss} respectively. 

To check the consistency of our results, we compare the $T_{\rm eff}$ values obtained from the three methods in Figure~\ref{fig:effective_temperature_relations_methods} against each other. The top figure is divided into two sub-panels, both of which are comparing $T_{\rm eff}$ derived with the isochrone fitting method against the values found with the IRFM (top--upper panel) and XPP (top--lower panel).~The lower panel compares the values of latter two methods against each other.~We only consider BSSs with an error in $T_{\rm eff}$ of less than $\pm300~{\rm K}$ in both components. The 1--to--1 relation is shown as a gray dashed line and is compared with the data. We observe very good agreement between the methods, albeit with small level of systematics. We fit a linear model using \texttt{LMFIT} Python package\footnote{\url{https://lmfit.github.io/lmfit-py/}} to these relations considering full error propagation in all values, which are shown as a dot-dashed green lines. We also estimated pearson r-coefficients for all these relations in a monte carlo approach, and the estimated values are 0.93$^{+0.0}_{-0.0}$ (top panel), 0.76$^{+0.03}_{-0.05}$ (middle panel) and 0.74$^{+0.04}_{-0.04}$ (bottom panel). The errors correspond to 16th and 84th percentile values with respect to the median value. The r-coefficients suggest that estimated $T_{\rm eff}$ from different methods are strongly correlated.  
We obtain the following linear relations in Kelvin units:

\begin{equation}
    {\rm T}_{\rm eff, IRFM} = (1.074 \pm 0.005) \times {\rm T}_{\rm eff, isochrone}  - (585 \pm 34) 
\end{equation}
\begin{equation}
    {\rm T}_{\rm eff, XPP} = (0.842 \pm 0.21) \times {\rm T}_{\rm eff, isochrone} + (1092 \pm 145)
\end{equation}
\begin{equation}
    {\rm T}_{\rm eff, XPP} = (0.735 \pm 0.026) \times {\rm T}_{\rm eff, IRFM} + (1870 \pm 173).
\end{equation}

From these results, we note that the $T_{\rm eff}$ extracted from isochrone fitting and those derived from the color--temperature relations (IRFM) appear most consistent with each other up to $\sim\!8000\,{\rm K}$, with little systematic offset towards higher IRFM temperature. A similar behavior is observed for the XPP versus isochrone method, where $T_{\rm eff}$ values are consistent within the scatter up to $\sim\!8000\,{\rm K}$. Above that temperature, a small systematic offset towards lower XPP temperatures is observed. The bottom panel of Figure~\ref{fig:effective_temperature_relations_methods} shows the XPP versus IRFM method comparison, which shows the largest scatter yet fairly consistent values for most BSSs, while the linear fit is pulled from 1--to--1 relation by some OC BSSs with temperatures above $\sim\!7500\,{\rm K}$.

From these comparisons, we estimate that the systematic uncertainty between various methods is of the order $\Delta T_{\rm eff}\simeq100$\,K. The IRFM and isochrone methods consider the metallicity of the cluster, while the XPP method is inferior for the reasons described in Section~\ref{sec:xpp}, which is reflected in the lower scatter and smaller systematic offset in the IRFM versus~isochrone method comparison (upper panel in Fig.~\ref{fig:effective_temperature_relations_methods}). In addition, XPP parameters are not available for stars at the MSTO and therefore we are not able to extract ${\rm T}_{\rm eff, MSTO}$ using XPP. We therefore consider the IRFM and isochrone method from now as the bona fide best $T_{\rm eff}$ estimates, keeping in mind the systematics described above.

\section{Astrophysical parameters for YSSs and RSSs} \label{YSS_RSS_parameters}

\begin{figure*}
    \includegraphics[width=0.75\textwidth]{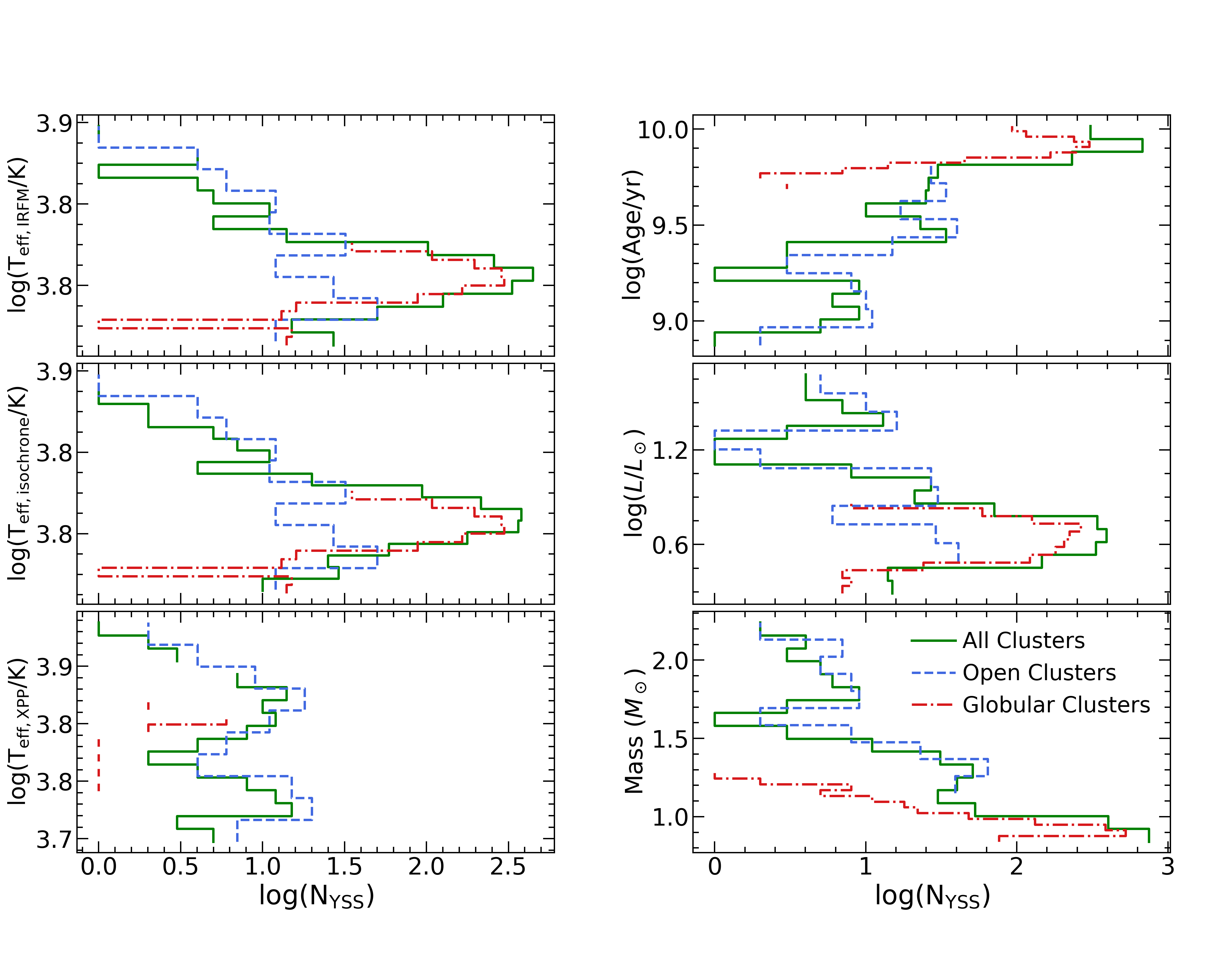}
    \caption{Same as Figure \ref{fig:histogram_teff_and_mass}, but for YSSs.}
    \label{fig:histogram_teff_and_mass_yss}
\end{figure*}

\begin{figure*}
    \includegraphics[width=0.75\textwidth]{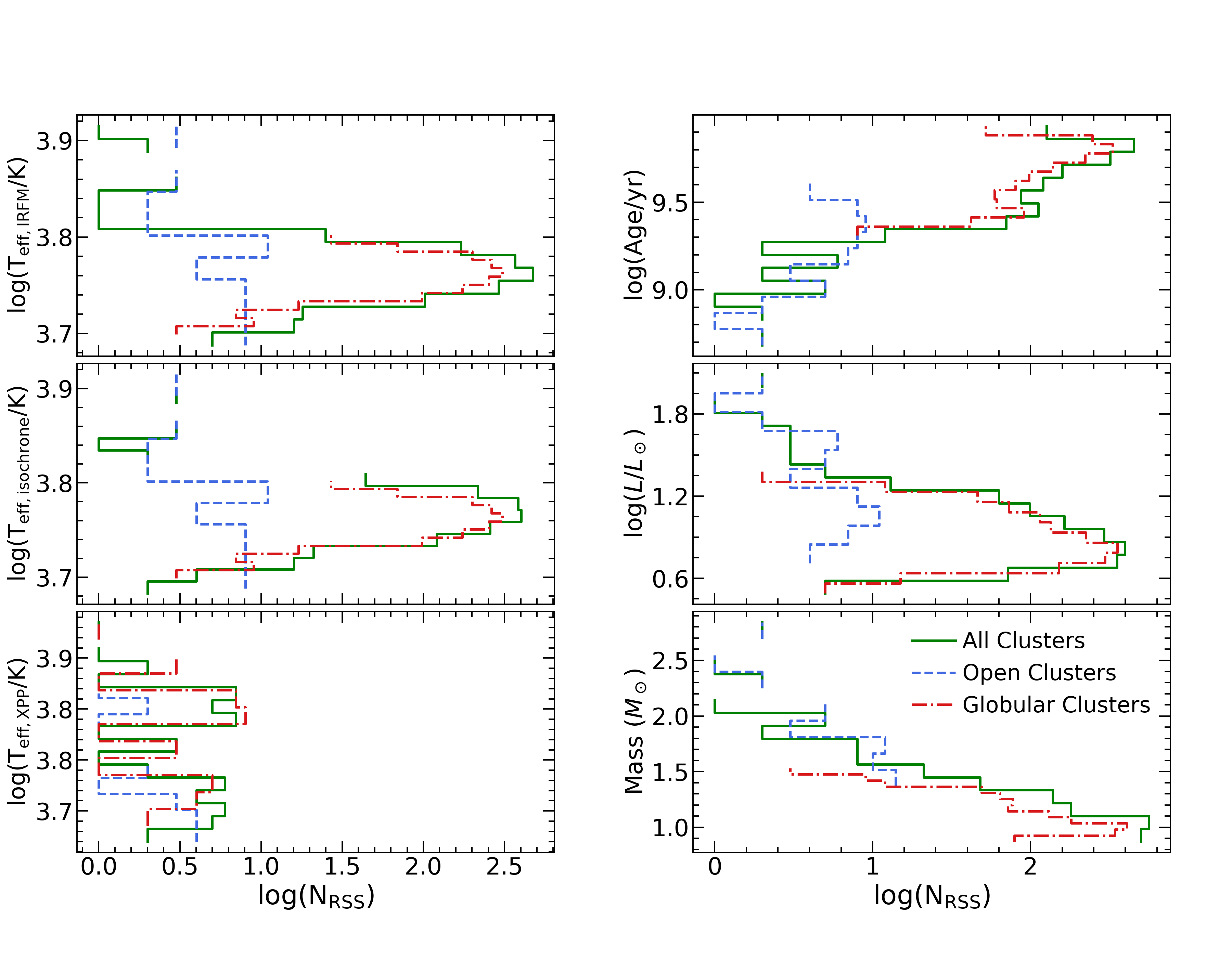}
    \caption{Same as Figure \ref{fig:histogram_teff_and_mass}, but for RSSs.}
    \label{fig:histogram_teff_and_mass_rss}
\end{figure*}

Similarly to Figure \ref{fig:histogram_teff_and_mass}, we display the extracted astrophysical parameters for stars classified as Yellow Straggler Stars (YSSs) and Red Straggler Stars (RSSs) in Figures \ref{fig:histogram_teff_and_mass_yss} and \ref{fig:histogram_teff_and_mass_rss}, respectively.

\begin{sidewaystable}
\section{Parameters of star clusters studied in this paper and the properties of BSSs detected in those clusters}

\tabcolsep2.50pt $ $
\caption{Parameters of star clusters are presented here.  }             
\label{tab:cluster_param}      
\centering                          
\begin{tabular}{c c c c c c c c  c c c  c c  }       
\hline\hline                 
cluster & cluster & RA & Dec & log(age/yr)  & Av & distance & [Fe/H] & T$_{eff,IRFM}$ & T$_{eff,isochrone}$ & N$_{BSS}$ & N$_{MS}$ & f$_{BSS/MS}$  \\
name & type & (deg) & (deg) &   & (mag) & (pc) & (dex) & (K) & (K) &  &  &   \\
(1) & (2) & (3) & (4) & (5) & (6) & (7) & (8) & (9) & (10) & (11) & (12) & (13) \\
\hline                        

NGC104 & GC & 6.175 & $-$72.024 &10.070$\pm$0.011&0.124&4367 & $-$0.65 &5903$\pm$92&5995&415&13691&0.06$\pm$0.002 \\
NGC288&GC&13.381&$-$26.529 &10.064$\pm$0.002&0.031&9000&$-$0.8 &6066$\pm$116&6173&150&--&-- \\
NGC362&GC&15.885&$-$70.781 &10.068$\pm$0.013&0.155&9090&$-$1.3 &6248$\pm$160&6368&137&1578&0.171$\pm$0.011 \\
\vdots & \vdots & \vdots & \vdots & \vdots & \vdots & \vdots & \vdots & \vdots & \vdots & \vdots & \vdots & \vdots \\
Berkeley39 & OC & 116.893 & $-$4.735 & 9.811 $\pm$ 0.029 & 0.394 & 4156 & 0.0 & 5932 $\pm$ 84 & 6038 & 22 & 264 & 0.14 $\pm$ 0.025 \\
Berkeley32 & OC & 104.738 & 6.379 & 9.686 $\pm$ 0.022 & 0.553 & 3321 & $-$0.3 & 6286 $\pm$ 125 & 6403 & 14 & 139 & 0.151 $\pm$ 0.035 \\
Collinder261 & OC & 189.802 & $-$68.492 & 9.898 $\pm$ 0.065 & 0.999 & 2706 & 0.1 & 5772 $\pm$ 72 & 5865 & 64 & 460 & 0.228 $\pm$ 0.025 \\
\vdots & \vdots & \vdots & \vdots & \vdots & \vdots & \vdots & \vdots & \vdots & \vdots & \vdots & \vdots & \vdots \\

\hline                                   
\end{tabular}
\tablefoot{The complete table is made available at the CDS. 
The list of columns represented here is:  (1) Name of the cluster; (2) whether it is a globular cluster (GC) or open cluster (OC); (3) and (4) represents central coordinates of the cluster, (5) age of the cluster in log scale, (6) extinction (A$_V$) towards the cluster, (7) distance to the cluster, (8) metallicity of the cluster, (9) mean effective temperature of BSS populations in that cluster using IRFM method, (10)  mean effective temperature of BSS populations in that cluster using isochrone method; (11) number of BSSs detected in that cluster; (12) number of main sequence (MS) stars in the luminosity range between the MS turn-off (MSTO) and MSTO+1 mag in that cluster; (13) fractional BSSs population to MS populations. }
\end{sidewaystable}
%

% --------
\begin{sidewaystable}
\tabcolsep1.750pt $ $
\caption{List of BSSs detected in this study and their properties. }\label{tab:bss}
\centering
\begin{tabular}{ccccc ccccc ccccc } 
\hline\hline   
 &  &  &   & \multicolumn{5}{c}{isochrone method} &  &  &  &  &  & \\
\cline{5-9}
cluster & type & Gaia source id &  T$_{eff,IRFM}$ & T$_{eff}$ & log(age/yr) & M & log(L/L$_\odot$) & logg & T$_{eff,xpp}$ & logg$_{xpp}$ & Me & Me$_{sequence}$ & ${BP}_{0}$ & $(BP-RP)_0$\\
 &  &  &  (K) & (K) &  & (M$_\odot$) &  &  & (K) &  &  &  & (mag) & (mag)\\
 (1) & (2) & (3) & (4) & (5) & (6) & (7) & (8) & (9) & (10) & (11) & (12) & (13) & (14) & (15)  \\
\hline
\hline

NGC2141 & OC & 3341815582307132416 & 6363$\pm$82 & 6510$\pm$59 & 9.222$\pm$0.015 & 1.70$\pm$0.01 & 1.040$\pm$0.006 & 3.833$\pm$0.010 & --  -- & --  -- & 0.17 & lower & 2.237 & 0.597 \\
NGC2141 & OC & 3341815582307133056 &  6391$\pm$84 & 6464$\pm$79 & 9.222$\pm$0.015 & 1.72$\pm$0.00 & 1.070$\pm$0.003 & 3.800$\pm$0.020 & --  -- & --  -- & 0.19 & lower & 2.176 & 0.589 \\
NGC2141 & OC & 3341815681086838272 &  6736$\pm$101 & 6855$\pm$95 & 9.087$\pm$0.077 & 1.56$\pm$0.01 & 0.800$\pm$0.005 & 4.122$\pm$0.025 & --  -- & --  -- & 0.08 & lower & 2.807 & 0.500 \\
\vdots & \vdots  & \vdots  &  \vdots  & \vdots  & \vdots  & \vdots  & \vdots  & \vdots  & \vdots  & \vdots  & \vdots  & \vdots  & \vdots  & \vdots  \\
NGC104 & GC & 4689572883825672192 & 6422$\pm$126 & 6567$\pm$46 & 9.740$\pm$0.027 & 1.01$\pm$0.01 & 0.550$\pm$0.004 & 4.119$\pm$0.015 & 6831$\pm$136 & 4.142$\pm$0.123 & 0.28 & upper & 3.538 & 0.566 \\
NGC104 & GC & 4689573433581464192 & 6262$\pm$109 & 6384$\pm$26 & 9.842$\pm$0.024 & 0.93$\pm$0.01 & 0.350$\pm$0.003 & 4.226$\pm$0.009 & 6896$\pm$166 & 4.216$\pm$0.2 & 0.18 & lower & 4.043 & 0.609 \\
NGC104 & GC & 4689574086416430464 & 6669$\pm$154 & 6838$\pm$35 & 9.505$\pm$0.035 & 1.08$\pm$0.01 & 0.490$\pm$0.002 & 4.271$\pm$0.009 & 6652$\pm$159 & 4.195$\pm$0.142 & 0.37 & upper & 3.651 & 0.505 \\
\vdots & \vdots  & \vdots  &  \vdots  & \vdots  & \vdots  & \vdots  & \vdots  & \vdots  & \vdots  & \vdots  & \vdots  & \vdots  & \vdots  & \vdots  \\
\hline
\hline

\hline
\end{tabular}
\tablefoot{The complete catalog is made available at the CDS. 
The list of columns presented here is (1) the name of the cluster to which the BSS belong; (2) whether the cluster is classified as an open cluster (OC) or globular cluster (GC); (3) source id of the BSS as per the Gaia DR3; (4) effective temperature of the BSS based on IRFM method; (5) -- (9) effective temperature, age, mass, luminosity and surface gravity of the BSS, respectively, estimated using isochrone method; (10) -- (11) effective temperature and $logg$ of the BSS estimated from Gaia-XPP spectrum. (12) fractional mass excess of the BSS based compared to MSTO mass; (13) location of the BSS based on mass excess, as mentioned in \autoref{fig:upper_lower_Me_classification}; (14)-(15) BP magnitude and (BP-RP) color in absolute scale. }
\end{sidewaystable}

\end{appendix}

\end{document}